\documentclass[preprintnumbers, onecolumn, floatfix, preprintnumbers, amsmath, amssymb, superscriptaddress]{revtex4}

\usepackage[colorlinks,linkcolor=red,urlcolor=blue,citecolor=blue]{hyperref}
\usepackage{amssymb, amsmath, bm, dcolumn, epsf, graphicx, latexsym, mathbbol, slashed}
\usepackage{mathrsfs}
\usepackage{comment}
\usepackage{color}
\usepackage{multirow}
\usepackage{array}
\usepackage{subcaption}
\usepackage{caption}
\begin{document}

\title{Schwarzschild Lensing From Geodesic Deviation}

\author{Zhao Li}
\email{lz111301@mail.ustc.edu.cn}
\affiliation{ Department of Astronomy, University of Science and Technology of China, Hefei, Anhui 230026, China}
\affiliation{ School of Astronomy and Space Science, University of Science and Technology of China, Hefei 230026, China}
\affiliation{ Department of Physics, Kyoto University, Kyoto 606-8502, Japan}
\author{Xiao Guo}

\affiliation{School of Fundamental Physics and Mathematical Sciences, Hangzhou Institute for Advanced Study, University of Chinese Academy of Sciences, No.1 Xiangshan Branch, Hangzhou 310024, China}
\author{Tan Liu}
\affiliation{School of Fundamental Physics and Mathematical Sciences, Hangzhou Institute for Advanced Study, University of Chinese Academy of Sciences, No.1 Xiangshan Branch, Hangzhou 310024, China}
\affiliation{University of Chinese Academy of Sciences, 100049/100190 Beijing, China}
\author{Tao Zhu}
\affiliation{Institute for Theoretical Physics and Cosmology, Zhejiang University of Technology, Hangzhou, 310032, China,\\
 United Center for Gravitational Wave Physics (UCGWP), Zhejiang University of Technology, Hangzhou, 310032, China}
\author{Wen Zhao}
\email{wzhao7@ustc.edu.cn}
\affiliation{ Department of Astronomy, University of Science and Technology of China, Hefei, Anhui 230026, China}
\affiliation{ School of Astronomy and Space Science, University of Science and Technology of China, Hefei 230026, China}

\begin{abstract}
We revisit the gravitational lensing of light or gravitational waves by a Schwarzschild black hole in geometric optics. 
Rather than focusing on a single massless particle, we investigate the collective behavior of a congruence of light or gravitational rays, described by the geodesic deviation equation (GDE). 
By projecting GDE onto the Newman-Penrose tetrad, we decouple the equation and find an analytical Dyson-like series solution in the weak deflection and thin lens limits. 
Using this solution, we analyze the evolution of the cross-sectional area and axis ratio. 
Finally, we reproduce the magnification and axis ratio of the lensing images up to second-order weak deflection approximation, addressing some previously overlooked corrections.
\end{abstract}

\maketitle

\section{\label{Introduction}Introduction}
Gravitational lensing is one of the most significant phenomena predicted by general relativity (GR), occurring when a massive object lies between a distant source and Earth \cite{Weinberg,MTW,Chandrasekhar1983}. The gravitational field bends light rays from the source, creating multiple images or producing arcs and rings \cite{Schneider1992}. One direct consequence is magnification, enabling us to observe objects that would otherwise be too distant and faint to detect at high redshift \cite{Wong2014}. Gravitational lensing serves as a powerful tool to probe the Universe and test gravity on the astrophysical scale \cite{XiaohuiLiu2022,Ezquiaga2021,QingqingWang2024,Refregier2003}. Since the lensing effect directly depends on the mass distribution of the lens object, including both visible and dark matter, it enables researchers to measure the dynamical mass of the galaxies and galaxy clusters, and to study the distribution of dark matter \cite{Koopmans2009}. 

Similar to light, gravitational waves (GWs) are lensed when they pass through massive objects \cite{Ezquiaga2021,Grespan2023,XiaoGuo2020,Takahashi2003,Mishra2021,Meena2019,LiangDai2017,Pagano2020}. Several studies have started searching for lensed GW signals in the current LIGO/Virgo/KAGRA catalog; however, no strong evidence for lensing imprints has been found \cite{Haris2018,Hannuksela2019,McIsaac2020,XiaohuiLiu2021,LIGO2021,LIGO2023,Lo2023}. As predicted, future third-generation ground-based GW observatories are expected to detect hundreds of lensed GW events, providing valuable insights into gravitational theory, cosmic structure, dark matter, and more \cite{LilanYang2021,ShaoqiHou2021,2022PhRvD.106b3018G}.

An ideal scenario arises when the typical wavelength of the lensed signal is much shorter than the background curvature scale. Through eikonal expansion, it is found that the behavior of light and GWs is analogous to that of massless particles in the lowest order, with their paths following the null geodesics \cite{Isaacson1968a,Isaacson1968b,ShaoqiHou2019}. Within this framework, the lensing process is described by the lens equation \cite{Keeton2005,Sereno2006}, given a specific surface mass density model for the lens object. The primary observational quantities of the lensing events consist of magnification and shear. In the conventional approach, these quantities are calculated using the Jacobian matrix, $\partial\bm{\beta}/\partial\bm{\theta}$ \cite{Keeton2005,Sereno2006}, where $\bm{\beta}$ and $\bm{\theta}$ represent the angular coordinates of the wave source and the lensed image. The lens equation $\bm{\beta}=\bm{\beta}(\bm{\theta})$ is derived by solving the null geodesic equation \cite{Sereno2006}.

In this work, we revisit the lensing process by a Schwarzschild black hole from a different perspective.The lensed signal consists not of a single photon/graviton, but of a congruence of light/gravitational rays \cite{Poisson2009}. During propagation, the geometry of the congruence is significantly affected by the tidal forces of the Schwarzschild lens \cite{Dolan2018a,Dolan2018b}. The geometry of the congruence is typically described by a deviation vector $\bm{\xi}$, which is the solution to the geodesic deviation equation (GDE) \cite{MTW}. Some references, such as \cite{Seitz1994}, also refer to $\bm{\xi}$ and the GDE as the Jacobi field and Jacobi equation. The solution for $\bm{\xi}$ at any point along the geodesics is generated by the Jacobi map, along with its initial values \cite{Seitz1994}. In this work, we study the evolution of geodesic congruence by solving the GDE, which is significantly different from the Jacobian matrix-based approach, where only the geodesic equation needs to be solved. In the weak deflection limit, where the impact distance is much larger than the gravitational radius of the lens \cite{Keeton2005,Sereno2006}, we find an analytical solution through the Dyson-like series expansion \cite{Boero2019}. Based on this, we analytically reproduce the magnification and axis ratio of the images. Furthermore, we refine the previous lens equation and its results, correcting the omission of several higher-order terms that were previously overlooked.

This paper is organized as follows. In Sec.\,\ref{sec:geodesic congruence}, we review the geodesic deviation equation (GDE) and project it onto the Newman-Penrose (NP) tetrad. In Sec.\,\ref{sec:solution}, we present the Dyson-like series solution to the GDE and discuss the physical interpretation of the optical scalars. In Sec.\,\ref{sec:lensing}, we explore Schwarzschild lensing, focusing on the evolution of the deviation vector and optical scalars, and most importantly, we reproduce the magnification and shear of the Schwarzschild lens. We summarize our findings in Sec.\,\ref{sec:conclusion}. Note that we do not consider the cosmological background or redshift in this work. Throughout the paper, we adopt geometric units in which $c=G=1$, where $c$ is the speed of light in a vacuum and $G$ is the gravitational constant.

\section{\label{sec:geodesic congruence}Geodesic congruence}

In this section, we investigate the behavior of the geodesic congruence \cite{Poisson2009}. The arm length of GW detectors and the aperture of optical telescopes are much smaller than the scale of lens objects. Therefore, the detected light and gravitational rays are very close to each other and approximately located within a common geodesic congruence.

For simplicity, we focus on two neighboring points, whose trajectories are denoted by
\begin{equation}
x^{\alpha}(\lambda,\mu^a),\quad\text{and}\quad
x^{\alpha}(\lambda+\delta\lambda,\mu^a+\delta\mu^a),
\end{equation}
where $\lambda$ is the affine parameter and $\mu^a$ is the $a$-th motion constant of a massless particle in the given background. The differences $\delta\lambda$ and $\delta\mu^a$ are assumed to be sufficiently small. The geodesic deviation vector is defined as the difference between such two trajectories, and can be expanded in terms of the small parameters $\delta\mu^a$ and $\delta\lambda$,
\begin{equation}
\xi^{\alpha}
\equiv x^{\alpha}(\lambda+\delta\lambda;\mu^a+\delta\mu^a)
-x^{\alpha}(\lambda;\mu^a)
=\frac{\partial x^{\alpha}}{\partial\lambda}\delta\lambda
+\sum_a\frac{\partial x^{\alpha}}{\partial\mu^a}\delta\mu^a
+\mathcal{O}(2).
\end{equation}
where $\mathcal{O}(2)$ denotes the second and higher order of $\delta\lambda$, $\xi^\alpha$, and $\delta\mu^a$. We denote the tangent vector of the geodesics as $k^{\mu}\equiv\partial x^{\mu}/\partial \lambda$, which satisfies the geodesic equation, $k^{\alpha}(\nabla_{\alpha}k^{\nu})=0$. The difference in wavevectors associated with each trajectory is then expanded in terms of the deviation vector, 
\begin{equation}
\label{delta-k-1}
\delta k^{\nu}=(\partial_{\alpha}k^{\nu})\xi^\alpha+\mathcal{O}(2).
\end{equation}
This difference can also be expanded in terms of small quantities $\delta\lambda$ and $\delta\mu^a$,
\begin{equation}
\label{delta-k-2}
\begin{aligned}
\delta k^{\nu}&=\sum_{a}\frac{\partial k^{\nu}}{\partial\mu^a}\delta\mu^a
+\frac{\partial k^{\nu}}{\partial\lambda}\delta\lambda
+\mathcal{O}(2)\\
&=\sum_{a}\frac{\partial^2x^{\nu}}{\partial\mu^a\partial\lambda}\delta\mu^a
+\frac{\partial^2x^{\nu}}{\partial\lambda^2}\delta\lambda
+\mathcal{O}(2)\\
&=\frac{\partial}{\partial\lambda}\left[\sum_{a}\frac{\partial x^{\nu}}{\partial\mu^a}\delta\mu^a
+\frac{\partial x^{\nu}}{\partial\lambda}\delta\lambda\right]
+\mathcal{O}(2)\\
&=\frac{\partial\xi^{\nu}}{\partial\lambda}
+\mathcal{O}(2)
=k^{\alpha}\partial_{\alpha}\xi^{\nu}
+\mathcal{O}(2),
\end{aligned}
\end{equation}
By combining Eqs.\,(\ref{delta-k-1}) and (\ref{delta-k-2}) up to the linear order of $\delta\lambda$ and $\delta\mu^\alpha$, we obtain \cite{Dolan2018a,Dolan2018b}
\begin{equation}
\label{GDV-1}
\xi^\alpha\nabla_{\alpha}k^{\nu}
=k^{\alpha}\nabla_{\alpha}\xi^{\nu},
\end{equation}
where $\nabla_{\alpha}$ is the covariant derivative operator compatible with the background metric. An equivalent interpretation to Eq.\,(\ref{GDV-1}) is that the Lie derivative of deviation vector along geodesic vanishes, i.e., $\mathcal{L}_{\bm{k}}\xi^{\nu}=k^{\alpha}(\nabla_{\alpha}\xi^{\nu})-(\nabla_{\alpha}k^{\nu})\xi^\alpha=k^{\alpha}(\partial_{\alpha}\xi^{\nu})-(\partial_{\alpha}k^{\nu})\xi^\alpha=0$. Multiplying $k_{\nu}$ on both sides of Eq.\,(\ref{GDV-1}), and combining the geodesic equation, one gets conservation law that 
\begin{equation}
k^{\alpha}\nabla_{\alpha}(k_{\nu}\xi^{\nu})=0,
\end{equation}
Setting the inner product $k_{\nu}\xi^{\nu}$ to zero means that the points $x^\alpha$ and $x^\alpha+\xi^\alpha$ lie on the same wavefront, with equal phases, as defined by $k_{\mu}\equiv-\nabla_{\mu}\Phi$ \cite{MTW}. This is demonstrated by
\begin{equation}
\Phi(x^\alpha+\xi^\alpha)\approx\Phi(x^\alpha)+(\partial_\alpha\Phi)\delta x^\alpha=\Phi(x^\alpha)-k_\alpha\xi^\alpha
=\Phi(x^\alpha).
\end{equation}
The GDE governs the evolution of the congruence, namely
\begin{equation}
\label{GDV-2}
\frac{D^2\xi^{\nu}}{D\lambda^2}
=-R^{\nu}_{\rho\alpha\beta}k^{\rho}k^{\beta}\xi^\alpha,
\end{equation}
where $D/D\lambda\equiv k^{\alpha}\nabla_{\alpha}$ and $R^{\nu}_{\rho\alpha\beta}$ is the background Riemann tensor. To solve Eq.\,(\ref{GDV-2}) more conveniently, we construct the NP tetrad along the null geodesics, with its four legs are denoted as \cite{NewmanPenrose1962,Chandrasekhar1983}
\begin{equation}
\bm{e}_{(a)}=\{\bm{k},\bm{n},\bm{m},\bar{\bm{m}}\}.
\end{equation}
The subscript $(a)$ denotes the tetrad indices. The first leg is the wavevector of the lensed waves, defined as the gradient of a scalar field $\Phi$. The second leg is in the spatially opposite direction. The third and fourth legs are orthogonal to the propagation direction of the waves and are complex conjugates of each other. The plane spanned by $\bm{m}$ and $\bar{\bm{m}}$ is generally referred to as the polarization plane, representing the oscillation direction of light or GWs. The orthogonality of the NP tetrad is expressed as $-\bm{k}\cdot\bm{n}=\bm{m}\cdot\bar{\bm{m}}=1$, with all other inner products vanish. 
Most importantly, all of the tetrad legs are required to be parallel-transported along the geodesics, $k^{\alpha}\nabla_{\alpha}e_{(a)}^{\mu}=0$. We project the deviation vector onto the NP tetrad as follows,
\begin{equation}
\label{GDVs-NP}
\xi^{\alpha}
=\xi^{(a)}e_{(a)}^{\alpha},\quad\text{with}\quad
\xi^{(a)}=\xi^{\alpha}e^{(a)}_{\alpha}.
\end{equation}
Setting $\xi^{\alpha}$ to satisfy the equal-phase condition, we obtain $\xi^{n}=k_{\alpha}\xi^{\alpha}=0$ \footnote{The superscript $n$ represents the components along $\bm{n}$-axis. A similar convention applies to the $m$ and $\bar{m}$ superscript in Eq.\,(\ref{GDE-1-NP-m}) and (\ref{GDE-2-NP-m}).}. Substituting Eq.\,(\ref{GDVs-NP}) into Eqs.\,(\ref{GDV-1},\,\ref{GDV-2}), we derive \cite{NewmanPenrose1962,Dolan2018a,Dolan2018b,Pineault1977,Seitz1994,Gallo2011,Dyer1977}
\begin{equation}
\label{GDE-1-NP-m}
\bm{\mathcal{D}}\xi^{m}
=\rho\xi^{m}+\bar{\sigma}\xi^{\bar{m}},
\end{equation}
and
\begin{equation}
\label{GDE-2-NP-m}
\bm{\mathcal{D}}^2\xi^{m}
=\bar{\Psi}_0\xi^{\bar{m}},
\end{equation}
for the $m$-components. The derivative operator is $\bm{\mathcal{D}}\equiv k^{\mu}\partial_{\mu}=d/d\lambda$. One obtains similar equations for $\bar{m}$-component by taking the complex conjugate of these two equations. In Eqs.\,(\ref{GDE-1-NP-m},\,\ref{GDE-2-NP-m}), the Greek letters $\rho$ and $\sigma$ represent the spin coefficients, defined as $\rho\equiv m^{\mu}(\nabla_{\mu}k_{\nu})\bar{m}^{\nu}$ and $\sigma\equiv m^{\mu}(\nabla_{\mu}k_{\nu})m^{\nu}$. These two scalars are also referred to as optical scalars, which describe the geometry and evolution of the geodesic congruence. The Weyl scalar $\Psi_{0}$ is defined as $\Psi_0\equiv-R_{\mu\nu\alpha\beta}k^{\mu}m^{\nu}k^{\alpha}m^{\beta}$.
We assumed the spacetime to be a vacuum throughout this work, in which case the Weyl tensor is identical to the Riemann tensor. Some details to derive Eqs.\,(\ref{GDE-1-NP-m},\,\ref{GDE-2-NP-m}) are shown in Appendix \ref{app:11_and_12}.

At the end of this section, we discuss the tetrad transformation of Eqs.\,(\ref{GDE-1-NP-m}) and (\ref{GDE-2-NP-m}). The three classes of tetrad rotation which preserve orthogonality are 
\begin{subequations}
\begin{align}
&\text{(1):}\quad\bm{k}\rightarrow\bm{k},\ \bm{m}\rightarrow\bm{m}+a\bm{k},\ 
\bm{n}\rightarrow\bm{n}+\bar{a}\bm{m}+a\bar{\bm{m}}+|a|^2\bm{k},
\label{rotation-1}\\
&\text{(2):}\quad\bm{n}\rightarrow\bm{n},\ \bm{m}\rightarrow\bm{m}+b\bm{n},\ \bm{k}\rightarrow\bm{k}+\bar{b}\bm{m}+b\bar{\bm{m}}+|b|^2\bm{n},
\label{rotation-2}\\
&\text{(3):}\quad\bm{k}\rightarrow A^{-1}\bm{k},\ 
\bm{n}\rightarrow A\bm{n},\ \bm{m}\rightarrow e^{i\psi}\bm{m}.
\label{rotation-3}
\end{align}
\end{subequations}
To ensure that the transformed tetrad satisfies the parallel-transported condition, $a$, $b$, $A$, and $\psi$ must be constant along the geodesics. In this work, the $\bm{k}$ is defined as the wavevector of photon/graviton, determined by solving the geodesic equation, such that we ignore the transformation involving $b$ and $A$. Under rotation \eqref{rotation-1}, $\xi^m$, $\Psi_0$, $\rho$, and $\sigma$ remain unchanged. Under rotation \eqref{rotation-3} with $A=1$, we have 
\begin{equation}
\label{tetrad-transformation}
\xi^m\rightarrow e^{-i\psi}\xi^m,\quad
\Psi_0\rightarrow e^{2i\psi}\Psi_0,\quad
\rho\rightarrow\rho,\quad
\sigma\rightarrow e^{2i\psi}\sigma.
\end{equation}
From transformation (\ref{tetrad-transformation}), we conclude that Eqs.\,(\ref{GDE-1-NP-m}) and (\ref{GDE-2-NP-m}) are tetrad-independent. This indicates that we can always choose the appropriate tetrad to make solving Eqs.\,(\ref{GDE-1-NP-m},\,\ref{GDE-2-NP-m}) as convenient as possible.

The simplest scenario occurs when $\Psi_0$ has a constant phase, allowing us to eliminate it by applying rotation (\ref{tetrad-transformation}), making $\Psi_0$ a real number, and simplifying Eq (\ref{GDE-2-NP-m}) as $\bm{\mathcal{D}}^2\xi^{m}
=\Psi_0\xi^{\bar{m}}$. By adding/subtracting this equation and its complex conjugate, and defining $\xi_{\pm}\equiv\xi^{m}\pm\xi^{\bar{m}}$, we get
\begin{equation}
\label{delta-pm-equation}
\bm{\mathcal{D}}^2\xi_{\pm}
=\pm\Psi_0\xi_{\pm},
\end{equation}
two decoupled equations about $\xi_{\pm}$. It is noted that the null tetrad leg $\bm{m}$ is related to a spacelike tetrad by $\bm{m}=(1/\sqrt{2})(\bm{e}_x+i\bm{e}_y)$. Therefore, $\xi_{\pm}$ are nothing but $\xi_x$ and $\xi_y$, two orthogonal components of geodesic deviation vector along basis $\bm{e}_x$ and $\bm{e}_y$.

In Sec \ref{sec:lensing}, we will demonstrate that when the NP tetrad is parallel-transported along the null geodesics, $\Psi_0$ in the Schwarzschild background retains a constant phase. Weyl scalar $\Psi_0$ becomes real when the basis vector $\bm{e}_x$ lies on the equatorial plane or equivalently the orbital plane of the particle. Therefore, the discussion in this article is restricted to Eq.\,\eqref{delta-pm-equation}.

\section{\label{sec:solution}Solution to deviation vector}
In this section, we present the solution to the geodesic vector from Eq.\,(\ref{delta-pm-equation}). By transforming these two equations into a set of first-order differential equations, we obtain the equivalent matrix form
\begin{equation}
\label{delta-pm-equation-matrix}
\bm{\mathcal{D}}\bm{\mathcal{Y}}
=\bm{\mathcal{A}}\bm{\mathcal{Y}},\quad\text{with}\quad
\bm{\mathcal{Y}}\equiv
\left(\begin{array}{c}
\xi_{\pm}\\
\bm{\mathcal{D}}\xi_{\pm}
\end{array}\right),\quad\text{and}\quad
\bm{\mathcal{A}}\equiv\left(\begin{array}{cc}
0 & 1\\ \pm\Psi_{0}& 0
\end{array}\right).
\end{equation}
In a typical lensing system, the impact parameter of the photon/graviton paths is much larger than the gravitational radius of the lens object. Meanwhile, the source and observer are far from the lens, and the background curvature approaches zero sufficiently at these two points. These two conditions are typically referred to as weak deflection and thin lens limits. We assume that the lensed wave is emitted by a point source and propagates freely near the source. Therefore, the initial conditions for the deviation vector can be set as $\xi_{\pm}=0$ and $\bm{\mathcal{D}}\xi_{\pm}=c_{\pm}$. The initial derivative is proportional to the open angle of the congruence near the wave source. In these two limits, the right-hand side of Eq.\,(\ref{delta-pm-equation-matrix}) is always perturbation, and its 0-th solution can be set as
\begin{equation}
\bm{\mathcal{Y}}_0=c_{\pm}\left(\begin{array}{c}
\lambda\\ 1
\end{array}\right).
\end{equation}
The Dyson-like series solution is then written as \cite{Boero2019,Gallo2011,Berry2024}
\begin{equation}
\bm{\mathcal{Y}}_{1}
=\int_0^{\lambda}d\lambda'\bm{\mathcal{A}}(\lambda')\bm{\mathcal{Y}}_{0}(\lambda'),
\end{equation}
\begin{equation}
\bm{\mathcal{Y}}_{2}
=\int_{0}^\lambda d\lambda'\bm{\mathcal{A}}(\lambda')\bm{\mathcal{Y}}_{1}(\lambda')
=\int_{0}^\lambda d\lambda'\int_{0}^{\lambda'}d\lambda''
\bm{\mathcal{A}}(\lambda')\bm{\mathcal{A}}(\lambda'')\bm{\mathcal{Y}}_{0}(\lambda''),
\end{equation}
$$\cdots,$$
\begin{equation}
\bm{\mathcal{Y}}_{n+1}
=\int_{0}^\lambda d\lambda'\bm{\mathcal{A}}(\lambda')\bm{\mathcal{Y}}_{n}(\lambda').
\end{equation}
The solution to $\xi_{\pm}$ is finally iteratively given by,
\begin{equation}
\label{GDV-Dyson}
\xi_{\pm}=c_{\pm}\tilde{\xi}_{\pm},\quad
\tilde{\xi}_{\pm}\equiv\sum_{n=0}^{\infty}(\pm1)^n\mathcal{I}_n,
\end{equation}
where the integrations $\mathcal{I}_n$ are
\begin{equation}
\label{In-int}
\mathcal{I}_n
=\int_0^\lambda d\lambda'\int_0^{\lambda'}d\lambda''\Psi_0(\lambda'')\mathcal{I}_{n-1}(\lambda'')
=\int_0^\lambda(\lambda-\lambda')\Psi_0(\lambda')\mathcal{I}_{n-1}(\lambda')d\lambda'
,\quad\text{with}\quad\mathcal{I}_0=\lambda.
\end{equation}
and $\tilde{\xi}_{\pm}$ are real. The $\mathcal{I}_n$ defined in Eq.\,(\ref{GDV-Dyson}) is just the first component of $\bm{\mathcal{Y}}_n$, except an overall factor $(-1)^nc_{\pm}$. Since $\xi_{+}$ is real and $\xi_{-}$ is purely imaginary, such that the integration constants $c_{+}$ and $c_{-}$ are real and purely imaginary, respectively. 

The GDE (\ref{GDE-2-NP-m}) is a second-order equation, and therefore, there should be two independent solutions. Supposing $\xi^{m}_1$ and $\xi^{\bar{m}}_2$ to be two independent solutions of deviation vector, each with different $c_{\pm}$. Therefore, from Eq.\,(\ref{GDE-1-NP-m}), the optical scalars are determined by \cite{Dolan2018a}
\begin{equation}
\label{rho-sigma-solution-form}
\left(\begin{array}{c}
\rho\\
\sigma
\end{array}\right)
=\left(\begin{array}{cc}
\xi^{\bar{m}}_1&\xi^{m}_1\\
\xi^{\bar{m}}_2&\xi^{m}_2
\end{array}\right)^{-1}
\left(\begin{array}{c}
\bm{\mathcal{D}}(\xi^{\bar{m}}_1)\\
\bm{\mathcal{D}}(\xi^{\bar{m}}_2)
\end{array}\right),
\end{equation}
and explicitly given by \cite{Frittelli2000},
\begin{equation}
\label{rho-sigma}
\rho=\bm{\mathcal{D}}\ln\sqrt{\left|\tilde{\xi}_{+}\tilde{\xi}_{-}\right|},\quad\text{and}\quad
\sigma=\bm{\mathcal{D}}\ln\sqrt{\left|\tilde{\xi}_{+}/\tilde{\xi}_{-}\right|},
\end{equation}
which is independent of the integration constant $c_{\pm}$.

Based on the previous statement, $\tilde{\xi}_{\pm}$ are identified as the geometric scale of geodesic congruence along the $\bm{e}_x$ and $\bm{e}_y$ directions. Thus, the product $\tilde{\xi}_{+}\tilde{\xi}_{-}$ is proportional to the cross-sectional area, and $\tilde{\xi}_{+}/\tilde{\xi}_{-}$ represents the axis ratio of the congruence. We can define such area and axis ratio as 
\begin{equation}
\label{area-ratio-definition}
\mathcal{A}\equiv\tilde{\xi}_{+}\tilde{\xi}_{-},\quad\text{and}\quad
\mathcal{R}\equiv\tilde{\xi}_{-}/\tilde{\xi}_{+},
\end{equation}
The corresponding evolution equations are derived as \cite{Dolan2018a,Shipley2019,Frittelli2000}
\begin{equation}
\label{evolution-area-axis-ratio}
\bm{\mathcal{D}}\mathcal{A}=2\rho\mathcal{A},\quad\text{and}\quad
\bm{\mathcal{D}}\mathcal{R}=-2\sigma\mathcal{R}.
\end{equation}
The physical meanings of the optical scalars are the relative rate of the cross-sectional area and the axis ratio of the null congruences.

\section{\label{sec:lensing}Application: Schwarzschild lensing}
\subsection{Null tetrad and Weyl scalars}
In the coordinates $\{t,r,\theta,\varphi\}$, the line element of Schwarzschild black hole of mass $M$ is \cite{Chandrasekhar1983,MTW,Weinberg}
\begin{equation}
ds^2=-f(r)dt^2
+f^{-1}(r)dr^2
+r^2(d\theta^2+\sin^2\theta d\varphi^2),\quad
f(r)\equiv1-\frac{2M}{r},
\end{equation}
and the tangent vector of the null geodesics equations are 
\begin{equation}
\label{wave-vector}
k^{\mu}=\left\{f^{-1}(r),
U_r,0,\frac{L}{r^2}\right\},
\end{equation}
Because of the spherical symmetry of the Schwarzschild background, the orbits of the massless test particle lie in the equatorial plane of the black hole, with $\theta=\pi/2$. $L$ is the angular momentum of the photon or graviton. The radial potential $U_r$ is defined as
\begin{equation}
U_{r}\equiv\pm\sqrt{1-\frac{L^2}{r^2}f(r)},
\end{equation}
where the overall sign is negative and positive when the particle moves toward or backward to the pericenter, respectively. In a general stationary spacetime, the spacelike basis $\bm{e}_y$ is constructed using the Newton gauge \cite{Brodutch2011,Vadapalli2024}, where the time component $e_y^0$ is set to $0$. Its spatial components $\vec{\mathbf{e}}_y$ are determined by defining a local $\vec{\mathbf{r}}$-axis along the free-fall acceleration, i.e., $\vec{\mathbf{r}}=(1,0,0)$, requiring the $\vec{\mathbf{e}}_y$ to be perpendicular to the 3-wavevector, $\vec{\mathbf{k}}$, and the local $\vec{\mathbf{r}}$-axis, i.e., $\vec{\mathbf{e}}_y\propto\vec{\mathbf{r}}\times\vec{\mathbf{k}}$. Finally, the basis vector $\bm{e}_{y}$ is 
\begin{equation}
\bm{e}_y=\left\{0,0,\frac{1}{r},0\right\}.
\end{equation}
And then another basis vector $\bm{e}_x$ is solved from the orthogonality, $\bm{e}_x\cdot\bm{e}_x=1$, $\bm{e}_x\cdot\bm{e}_y=0$, and $\bm{e}_x\cdot\bm{k}=0$,
\begin{equation}
\bm{e}_x=\frac{1}{E}\left\{0,-\frac{L}{r}f(r),0,\frac{U_{r}}{r}\right\}.
\end{equation}
The transverse null tetrad vector $\bm{m}_0$ is defined as 
\begin{equation}
\label{tetrad-m0}
\bm{m}_0=\frac{1}{\sqrt{2}}(\bm{e}_x+i\bm{e}_y)
=\frac{1}{\sqrt{2}}\left\{0,-\frac{L}{E}\frac{f(r)}{r},\frac{i}{r},\frac{1}{E}\frac{U_{r}}{r}\right\},
\end{equation}
and $\bm{n}_0$ is obtained by solving orthogonality $\bm{k}_0\cdot\bm{n}_0=-1$, $\bm{n}_0\cdot\bm{n}_0=0$, and $\bm{n}_0\cdot\bm{m}_0=0$, presented as
\begin{equation}
\label{tetrad-n0}
\bm{n}_0=\frac{1}{2E}\left\{1,-\frac{f(r)}{E}U_r,0,-\frac{L}{E}\frac{f(r)}{r^2}\right\}.
\end{equation}
The null tetrad (\ref{wave-vector}), (\ref{tetrad-n0}), and (\ref{tetrad-m0}) satisfies the orthogonality condition but does not satisfy the parallel-transported condition. Using Rotation \eqref{rotation-1}, we can construct the parallel-transported tetrad as 
\begin{equation}
\label{tetrad}
\bm{k}=\bm{k},\quad
\bm{m}=\bm{m}_0+\alpha\bm{k},\quad
\bm{n}=\bm{n}_0+\alpha(\bm{m}_0+\bar{\bm{m}}_0)+\alpha^2\bm{k},
\end{equation}
where $\alpha$ is function about $r$ and determined by
\begin{equation}
\frac{d\alpha}{dr}=\frac{ML}{\sqrt{2}Er^3}U_r^{-1}.
\end{equation}
Under tetrad (\ref{tetrad}), we calculate $\Psi_0$ as
\begin{equation}
\Psi_0=\frac{3ML^2}{r^5}.
\end{equation}
Note that $\Psi_0$ is a spin-2 quantity and its value depends on the choice of the tetrad. In this work, we have chosen $\bm{e}_y$ to lie on the equatorial plane and $\bm{e}_x$ perpendicular to the equatorial plane, $\Psi_0$ is a real number, ensuring that we can apply the framework in Sec \ref{sec:solution} to solve GDE in Schwarzschild background. If we rotate $\bm{m}$, or equivalently $\vec{\mathbf{e}}_x$ and $\vec{\mathbf{e}}_y$, by an angle $\psi$ around the wave vector $\vec{\mathbf{k}}$, we have $\Psi_0\rightarrow\Psi_0e^{2i\psi}$, producing a constant phase of $2\psi$. However, the GDE is covariant under such tetrad transformations, without influence on the subsequent derivation.

\subsection{Analytical solution under the weak deflection limit}
We use the Dyson-like series to re-express the solution to the geodesic deviation vector and optical scalars. This work analytically integrates these integrations under weak deflection and thin-lens assumptions. In these limits, the angular momentum $L$ (with mass dimension) is much larger than the Schwarzschild radius, therefore, the impact parameter $b\equiv L/M\gg1$. The distances to the emitter and observer are much larger than the impact distance, i.e. $r_s, r_o\gg L$. The solutions to the null geodesic equation can be found in Ref.\,\cite{Chandrasekhar1983} and Appendix \ref{app:lensing}.

To perform the integration (\ref{In-int}), it is convenient to express the affine parameter in terms of a new parameter $x$, defined as $x\equiv u/u_m$, with $u\equiv M/r$ and $u_m\equiv M/r_m$, where $r_m$ is the pericenter radius of the photon/graviton trajectory. The solution to affine parameter $\lambda=\lambda(x)$ is given in Appendix \ref{app:lambda-sol}, where two distinct forms of solution exist in the intervals $[\lambda_s,\lambda_m]$ and $[\lambda_m,\lambda_o]$, denoted by $\lambda_{\rm I}(x)$ and $\lambda_{\rm II}(x)$. Here, $\lambda_s,\lambda_m,\lambda_o$, represent the affine parameter $\lambda$ evaluated at the source, pericenter, and observer positions, respectively. As a result, the integration takes on different forms in these two intervals. When $\lambda_s<\lambda<\lambda_m$, the integral is expressed as
\begin{equation}
\label{Dyson-int-x-1}
\mathcal{I}_n(x)=\int_{x_s}^{x}[\lambda_{\rm I}(x)-\lambda_{\rm I}(x')]\Psi_0[\lambda_{\rm I}(x')]\mathcal{I}_{n-1}[\lambda_{\rm I}(x')]\frac{d\lambda_{\rm I}(x)}{dx'}dx',\quad(n=1,2,3,\cdots).
\end{equation}
When $\lambda_m<\lambda<\lambda_o$, the integral is split into two parts:
\begin{equation}
\label{Dyson-int-x-2}
\begin{aligned}
\mathcal{I}_n(x)&=\int_{x_s}^{1}[\lambda_{\rm II}(x)-\lambda_{\rm I}(x')]\Psi_0[\lambda_{\rm I}(x')]\mathcal{I}_{n-1}[\lambda_{\rm I}(x')]\frac{d\lambda_{\rm I}(x)}{dx'}dx'\\
&\qquad+\int_{1}^{x}[\lambda_{\rm II}(x)-\lambda_{\rm II}(x')]\Psi_0[\lambda_{\rm II}(x')]\mathcal{I}_{n-1}[\lambda_{\rm II}(x')]\frac{d\lambda_{\rm II}(x)}{dx'}dx',\quad(n=1,2,3,\cdots).
\end{aligned}
\end{equation}
The analytical expressions for the integrations $\mathcal{I}_n$ are presented in Appendix \ref{app:analytical-In}. Due to the complexity of these expressions, we performed numerical calculations to better illustrate their behavior.

\begin{figure}[ht]
\begin{subfigure}{0.48\textwidth}
\includegraphics[width=1.0\textwidth]{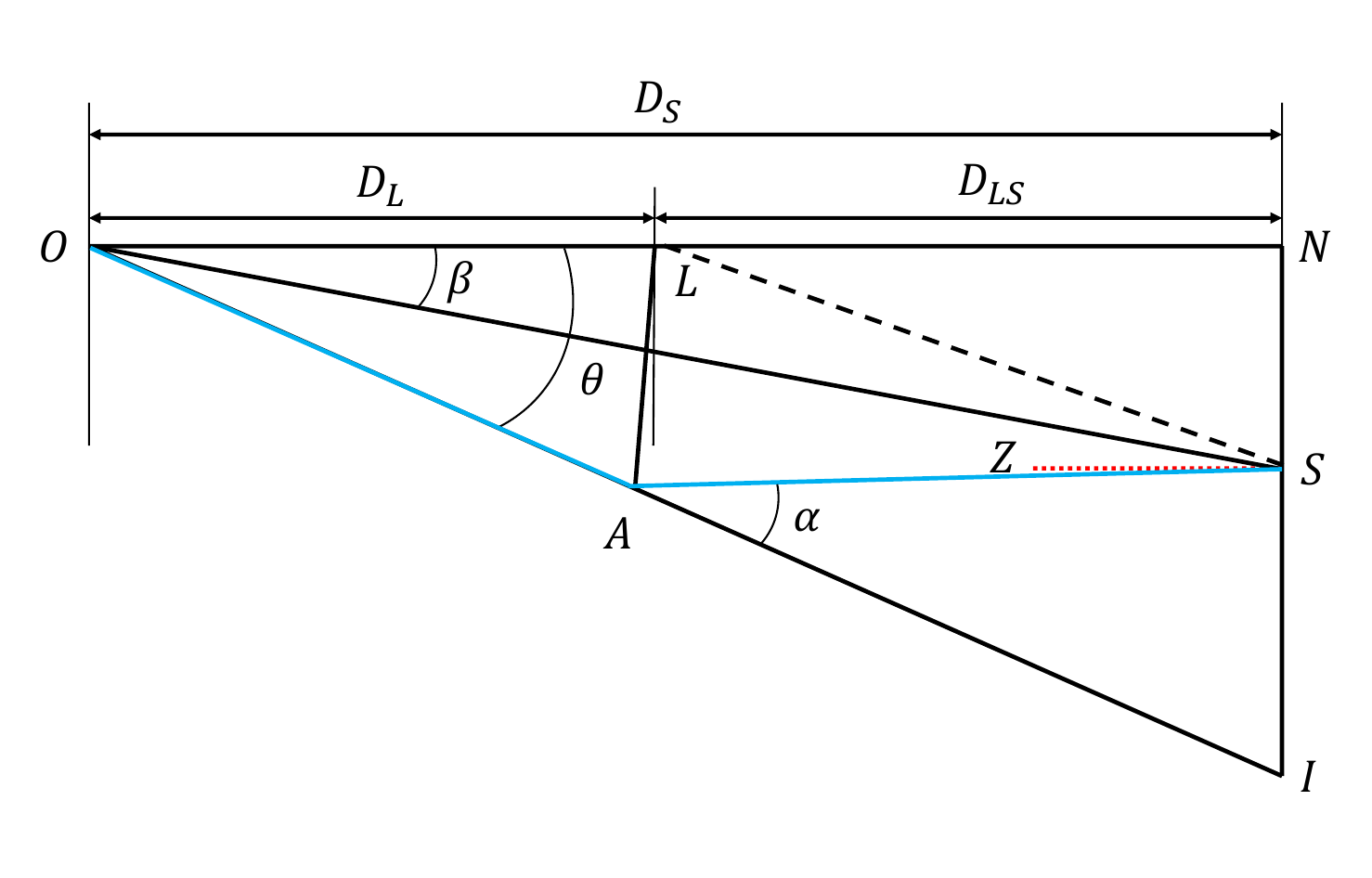}
\caption{Image 1}
\label{fig:img1}
\end{subfigure}
\begin{subfigure}{0.48\textwidth}
\includegraphics[width=1.0\textwidth]{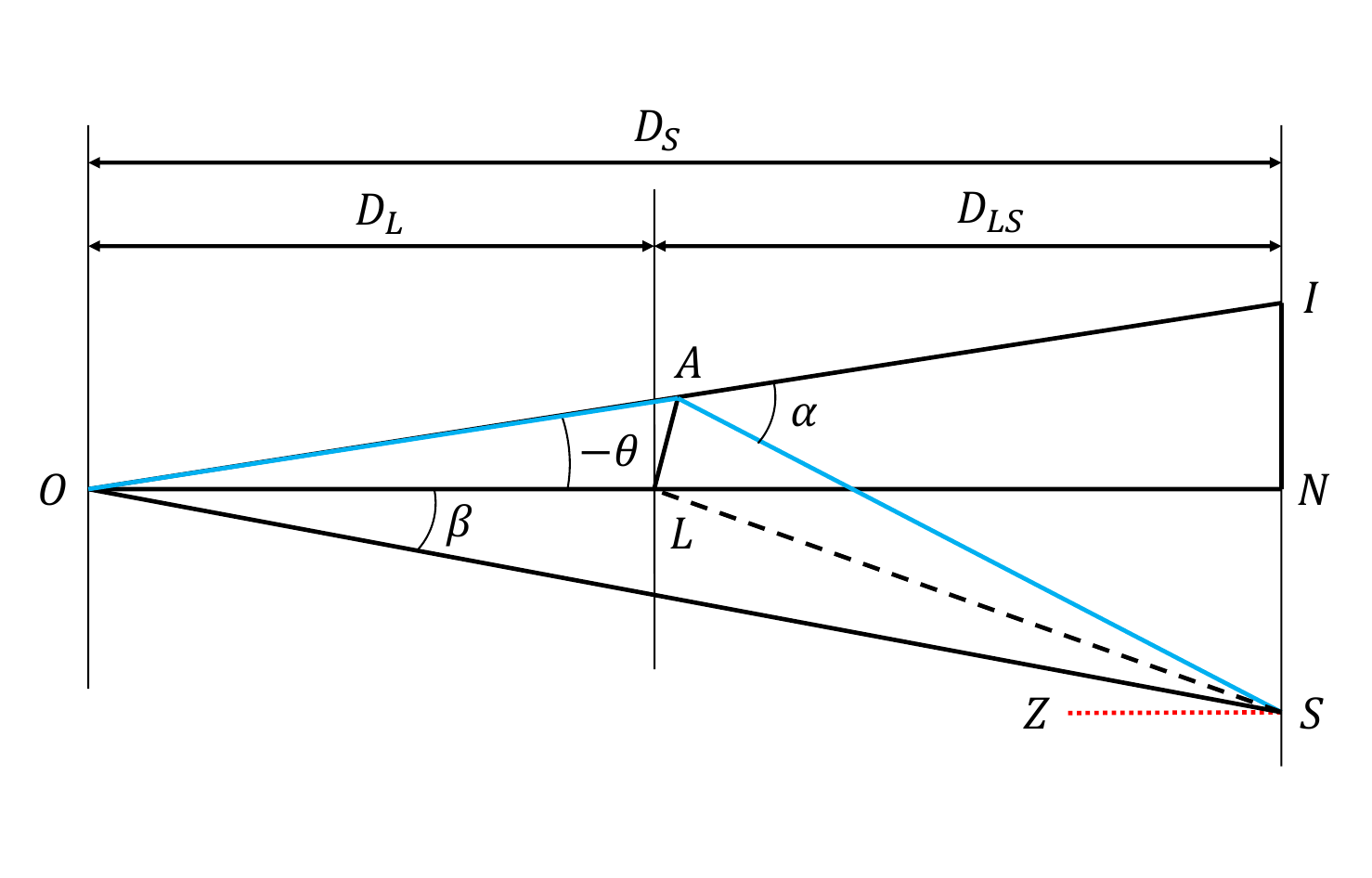}
\caption{Image 2}
\label{fig:img2}
\end{subfigure}
\caption{Geometrical configuration of gravitational lensing system (left: for image 1, right: for image 2). The points O, L, and S represent the observer, the lens object, and the wave source. The straight line ON represents the optical axis. The planes passing through the lens and source, both perpendicular to the optical axis are referred to as the lens and source plane. The distances from the observer to the lens and source planes are denoted by $D_{L}$ and $D_{S}$, respectively, with $D_{LS}=D_{S}-D_{L}$ representing the distance between the lens and the source. The deflection angle is denoted by $\alpha$, and the angular coordinates of the source and image are $\beta$ and $\theta$. Point Z is an auxiliary point, and the line SZ is perpendicular to the source plane.}
\label{fig:lensing}
\end{figure}

The typical configuration of gravitational lensing under the weak deflection and thin lens approximations is depicted in Fig.\,\ref{fig:lensing}. The two panels correspond to the two possible photon/graviton paths, forming two images in the geometric optics. For simplicity, we denote the left panel as image 1 and the right panel as image 2. In Fig.\,\ref{fig:lensing},  O, L, and S represent the observer, lens object, and wave source, respectively. The light blue line roughly indicates the photon/graviton path, and I is the resulting image on the source plane. Correspondingly, the line OS represents the free-propagation path. And $\alpha$, $\beta$, and $\theta$ are the deflection angle, the angular coordinate of the source and image. These quantities are related by lens equation, discussed in Appendix \ref{app:lensing}. The symbols $D_{L}$, $D_{S}$, and $D_{LS}$ represent the distances between the observer, the lens plane, and the source plane. In our numerical calculation, we set $D_{L}=780M$ and $D_{LS}=750M$. The well-known Einstein angle is defined by \cite{Schneider1992}
\begin{equation}
\label{Einstein-angle}
\theta_E\equiv \sqrt{4M\frac{D_{LS}}{D_SD_L}},
\end{equation}
which, for our setup, evaluates to approximately $0.0501381\,\mathrm{rad}\approx2.8727^\circ $. This small value is consistent with the weak deflection assumption. Given the distances $D_{L}$ and $D_{LS}$ (in units of lens mass), the only free parameter is the angular coordinate of the wave source, $\beta$. For convenience, we introduce the normalized coordinate $\beta_{0}\equiv\beta/\theta_{E}$ and set $\beta_0=\{0.2,0.4,0.6,0.8\}$. The corresponding image position, deflection angle, and impact parameter are listed in Table \ref{tab:lensing-parameters}. In the weak deflection limit, these quantities satisfy $\beta,\theta,\alpha,u_m\ll1$ and $b\gg1$.

\renewcommand\arraystretch{1.4}
\begin{table}[h]
\centering
\caption{We list the example values of the lensing parameters involved in our discussion. With $D_{L}=780M$ and $D_{LS}=750M$ fixed, the system is entirely determined by source position. The first two columns show the (normalized) source coordinates. The remaining columns present the normalized image coordinates, deflection angle, impact parameter, magnification, and axis ratio, for the two images produced by the Schwarzschild lensing. For each value of $\beta_0$, the first row corresponds to the value for image 1, while the second row corresponds to image 2.}
\setlength{\tabcolsep}{2mm}{
\begin{tabular}{cccccccccc}
\hline\hline
$\beta_0\equiv\beta/\theta_E$ & 
$\beta\,({\mathrm{rad}})$ & 
$\theta_0\equiv\theta/\theta_{E}$ & 
$\alpha\,({\mathrm{rad}})$ & 
$b$ & 
$u_m$ &
$\lambda_c(M)$ &
$\mu$ &
$\mathcal{R}$\\ 
\hline
\multirow{2}{*}{$0.2$} & 
\multirow{2}{*}{$0.0100276$} & 
$+1.1049880$ & 
$0.0961826$ & 
$44.57491$ & 
$0.0229677$ &
/ &
$+3.0300350$ &
$+0.0932811$\\
\cline{3-9}
& & 
$-0.9049876$ & 
$0.1174096$ & 
$37.06395$ & 
$0.0277617$ &
$1295.4090$ &
$-2.0459700$ &
$-0.0910933$\\
\hline
\multirow{2}{*}{$0.4$} & 
\multirow{2}{*}{$0.0200552$} & 
$+1.2198040$ & 
$0.0870816$ & 
$48.91776$ & 
$0.0208831$ &
/ &
$+1.8169000$ &
$+0.1857575$ \\
\cline{3-9}
& & 
$-0.8198039$ & 
$0.1294972$ & 
$33.88811$ &
$0.0304500$ &
$1152.5960$ &
$-0.8320377$ &
$-0.1772210$ \\
\hline
\multirow{2}{*}{$0.6$} & 
\multirow{2}{*}{$0.0300828$} & 
$+1.3440310$ & 
$0.0789617$ & 
$53.63810$ & 
$0.0190082$ &
/ &
$+1.4357270$ &
$+0.2746534$\\
\cline{3-9}
& & 
$-0.7440307$ & 
$0.1424925$ & 
$31.07548$ & 
$0.0333071$ &
$1058.9230$ &
$-0.4496574$ &
$-0.2561892$\\
\hline
\multirow{2}{*}{$0.8$} & 
\multirow{2}{*}{$0.0401105$} & 
$+1.4770330$ & 
$0.0717680$ & 
$58.71282$ & 
$0.0173352$ &
/ &
$+1.2607480$ &
$+0.3577646$\\
\cline{3-9}
& & 
$-0.6770330$ & 
$0.1563105$ & 
$28.59841$ & 
$0.0363085$ &
$994.5211$ &
$-0.2732032$ &
$-0.3265907$\\
\hline\hline
\end{tabular}}

\label{tab:lensing-parameters}
\end{table}

\begin{figure}[h]
\centering
\includegraphics[width=1.0\linewidth]{"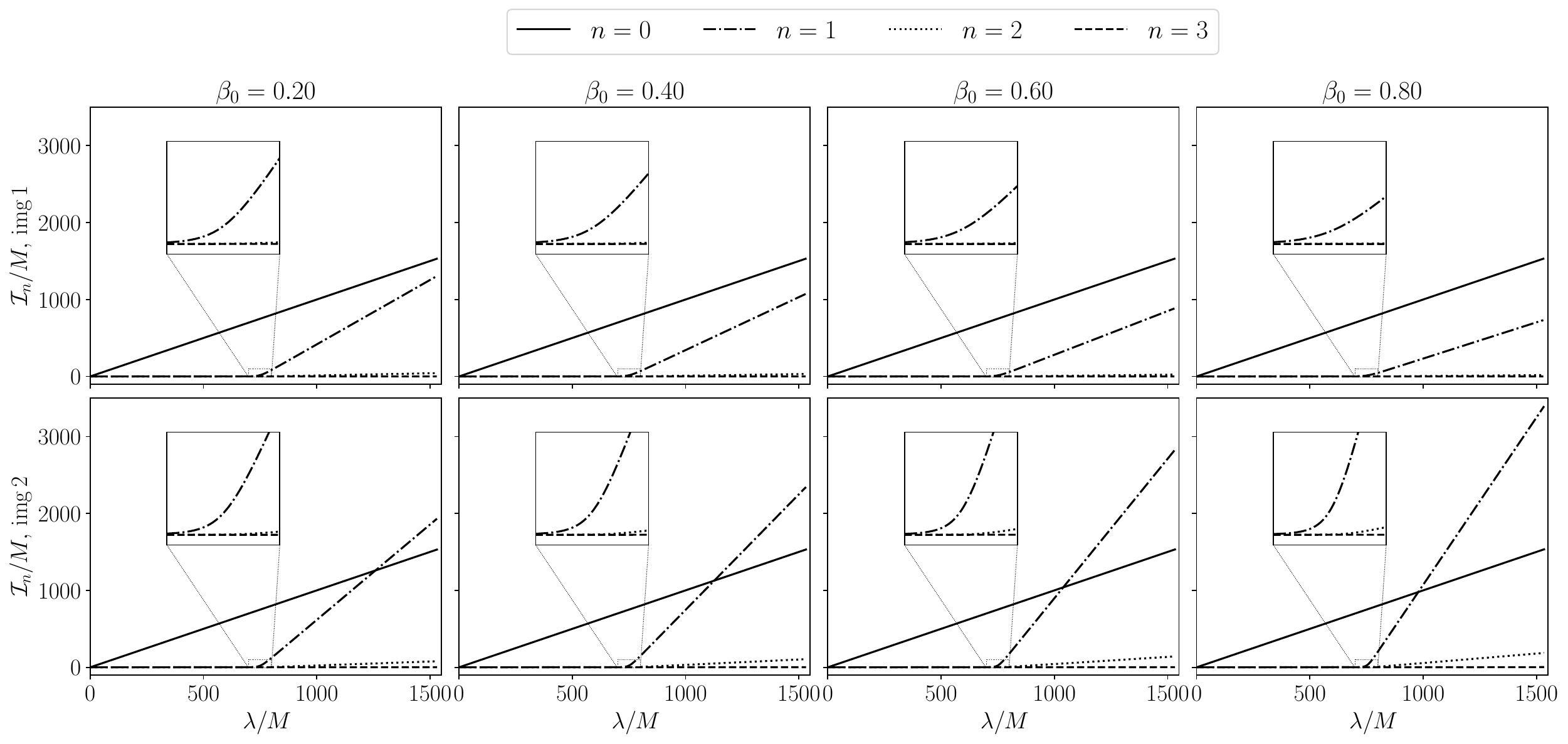"}
\caption{The evolution of the first four terms in the Dyson-like series (\ref{GDV-Dyson}) along the geodesic. These curves are plotted based on the analytical solution shown in Eqs.\,(\ref{lambda_1},\,\ref{lambda_2},\,\ref{I1-analytic},\,\ref{I2-analytic},\,\ref{I3-analytic}).}
\label{fig:Dyson_Series}
\end{figure}

In Fig.\,\ref{fig:Dyson_Series}, we show the evolution of the first four terms in the Dyson-like series given in Eq.\,(\ref{In-int}). The first two terms $\mathcal{I}_0=\lambda$, represented as a straight line with slope $1$, and $\mathcal{I}_1$ are the dominant contributions to the geodesic deviation vector. An obvious turning point is observed at $\lambda/M\sim750$, which indicates that the external gravitational field deflects the photon/graviton path at slightly different angles. The subsequent terms, $\mathcal{I}_2$ and $\mathcal{I}_3$, are nearly zero, representing high-order weak deflection correction. This suggests that the Dyson-like series converges within the weak-deflection limit.

\begin{figure}[ht]
\centering
\includegraphics[width=1.0\linewidth]{"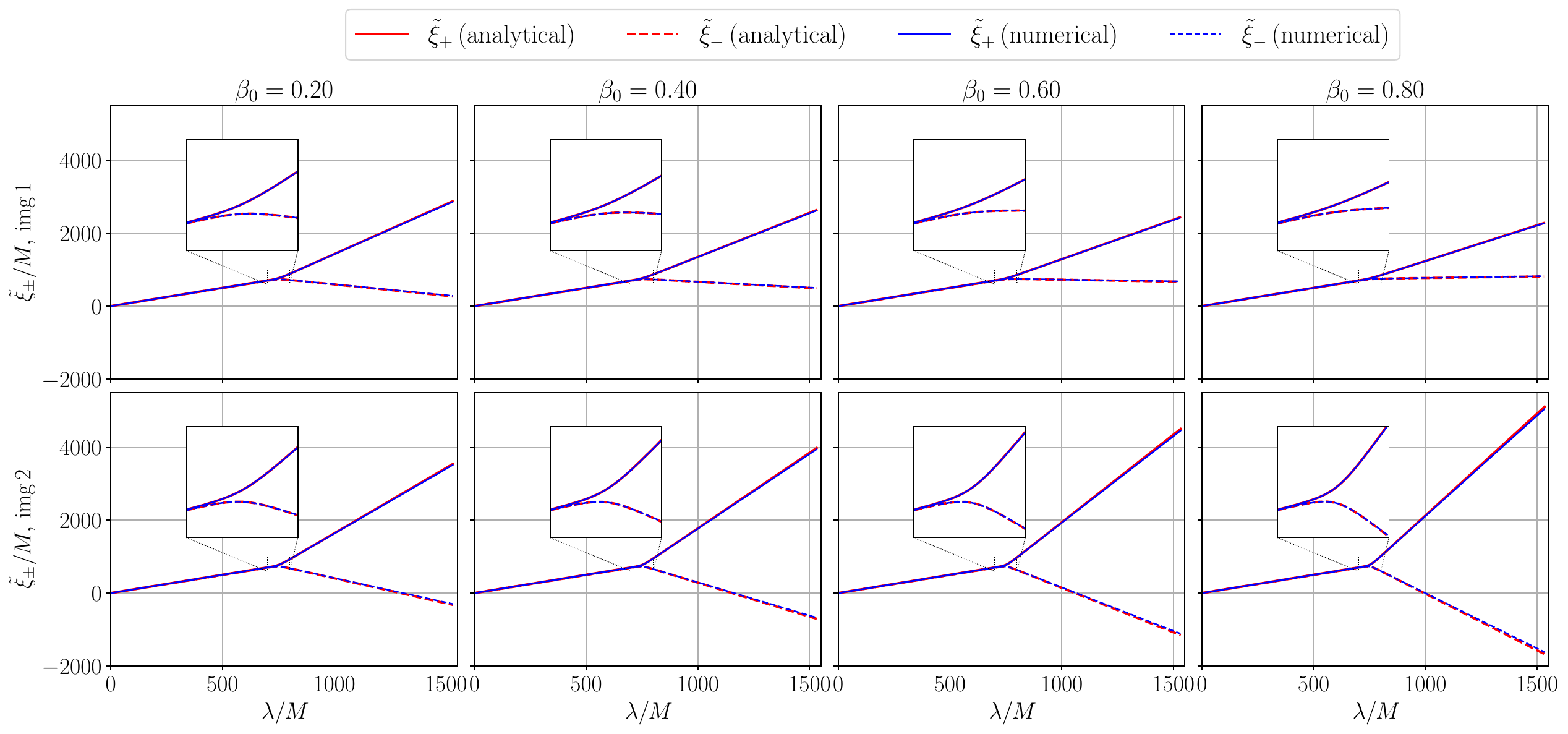"}
\caption{The evolution of the transversal components of the geodesic deviation vector, $\tilde{\xi}_{\pm}$ from different source positions and images. The red curves show the evaluated results based on the analytical expressions from the Dyson-like series (\ref{In-int}). The blue curves are obtained by numerically solving the GDE (\ref{delta-pm-equation}), together with the radial geodesic equation in Eq.\,(\ref{wave-vector}).}
\label{fig:geodesic_deviation_vector}
\end{figure}

Fig.\,\ref{fig:geodesic_deviation_vector} illustrates the evolution of the transversal components of the geodesic deviation vector along the null geodesics. Before the photon/graviton enters the external gravitational field, the evolutions of $\tilde{\xi}_{+}$ and $\tilde{\xi}_{-}$ are nearly identical. However, after exiting the gravitational field, their slopes differ significantly. As we mentioned before, the component $\tilde{\xi}_+$ lies on the equatorial plane of the Schwarzschild black hole, while $\tilde{\xi}_-$ is perpendicular to it. Due to the tidal forces, $\tilde{\xi}_+$ continually increases, with the rate of increase becoming larger after leaving the lens object than before entering. In contrast, the growth of $\tilde{\xi}_-$ is suppressed. For relatively large impact parameters $b$, the tidal force is strong enough to cause $\tilde{\xi}_-$ to decrease rapidly, as shown in the second row of Fig.\,\ref{fig:geodesic_deviation_vector}. Here, $\tilde{\xi}_-$ reaches zero, marking the moment where the cross-sectional area and axis ratio approaches $0$. This is known as the caustic point, denoted by $\lambda_c$, where $\tilde{\xi}_-(\lambda_c)=0$. At this point, the lensing image is inverted (negative parity). On the other hand, for the smaller impact parameters, no caustic point is observed, and $\tilde{\xi}_-$ remains positive throughout the entire path, ultimately resulting in an upright (positive parity) image. In addition to the approximate analytical solution, we also present the numerical solution to Eq.\,(\ref{delta-pm-equation}), with initial conditions $\xi_{\pm}(\lambda=0)=0$ and $\bm{\mathcal{D}}\xi_{\pm}(\lambda=0)=1$. These results, also shown in Fig.\,\ref{fig:geodesic_deviation_vector}, validate that the Dyson-like series solution is accurate under the weak deflection and thin lens approximations. 

\begin{figure}[ht]
\centering
\includegraphics[width=0.8\linewidth]{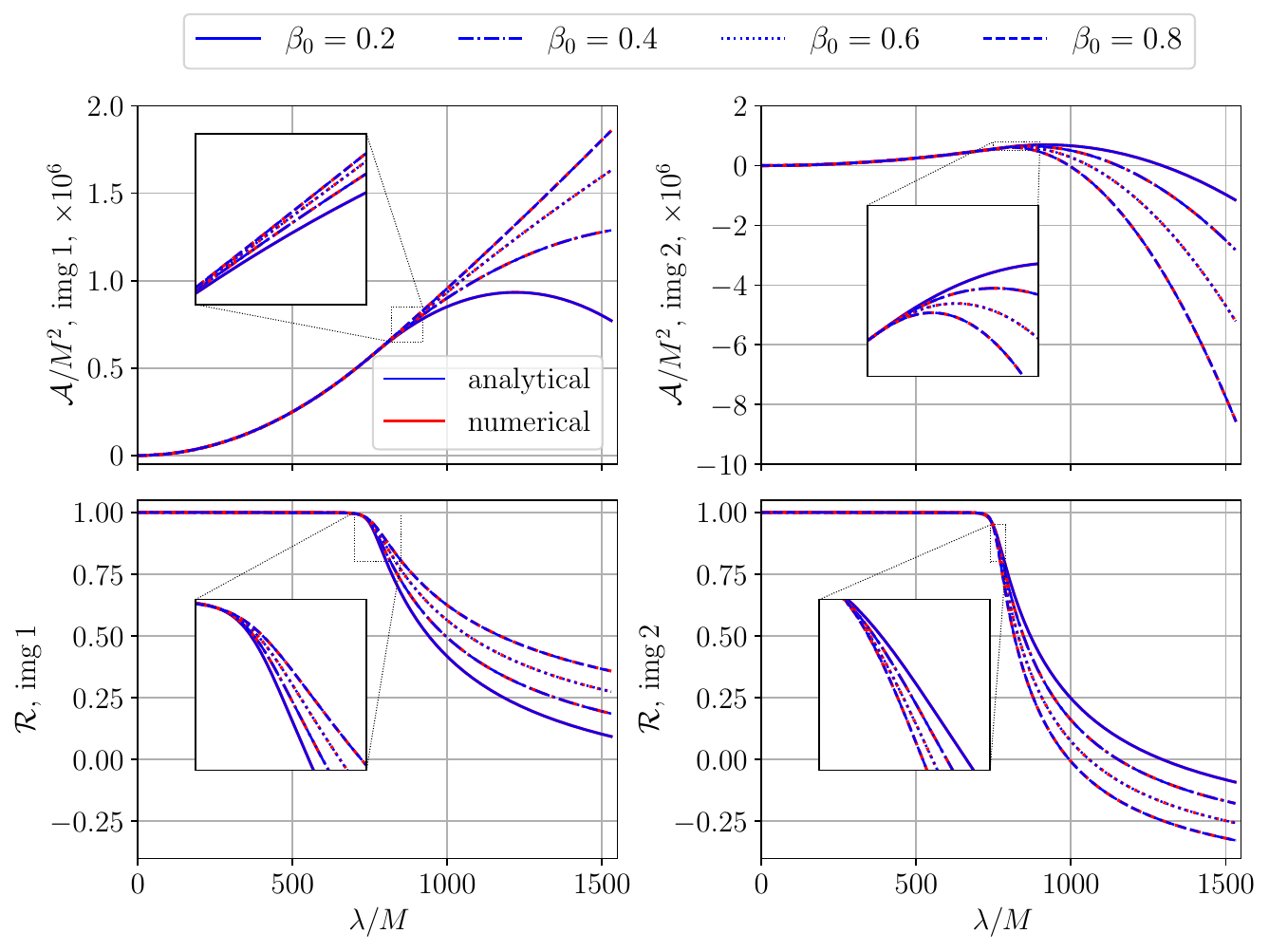}
\caption{The evolution of the area and axis ratio of the geodesic congruence. Same with Fig.\,\ref{fig:geodesic_deviation_vector}, the blue and red curves are obtained from the analytical solutions and corresponding numerical solutions to Eq.\,(\ref{delta-pm-equation}).}
\label{fig:magnification_and_axis_ratio}
\end{figure}

\begin{figure}[ht]
\centering
\includegraphics[width=0.8\linewidth]{"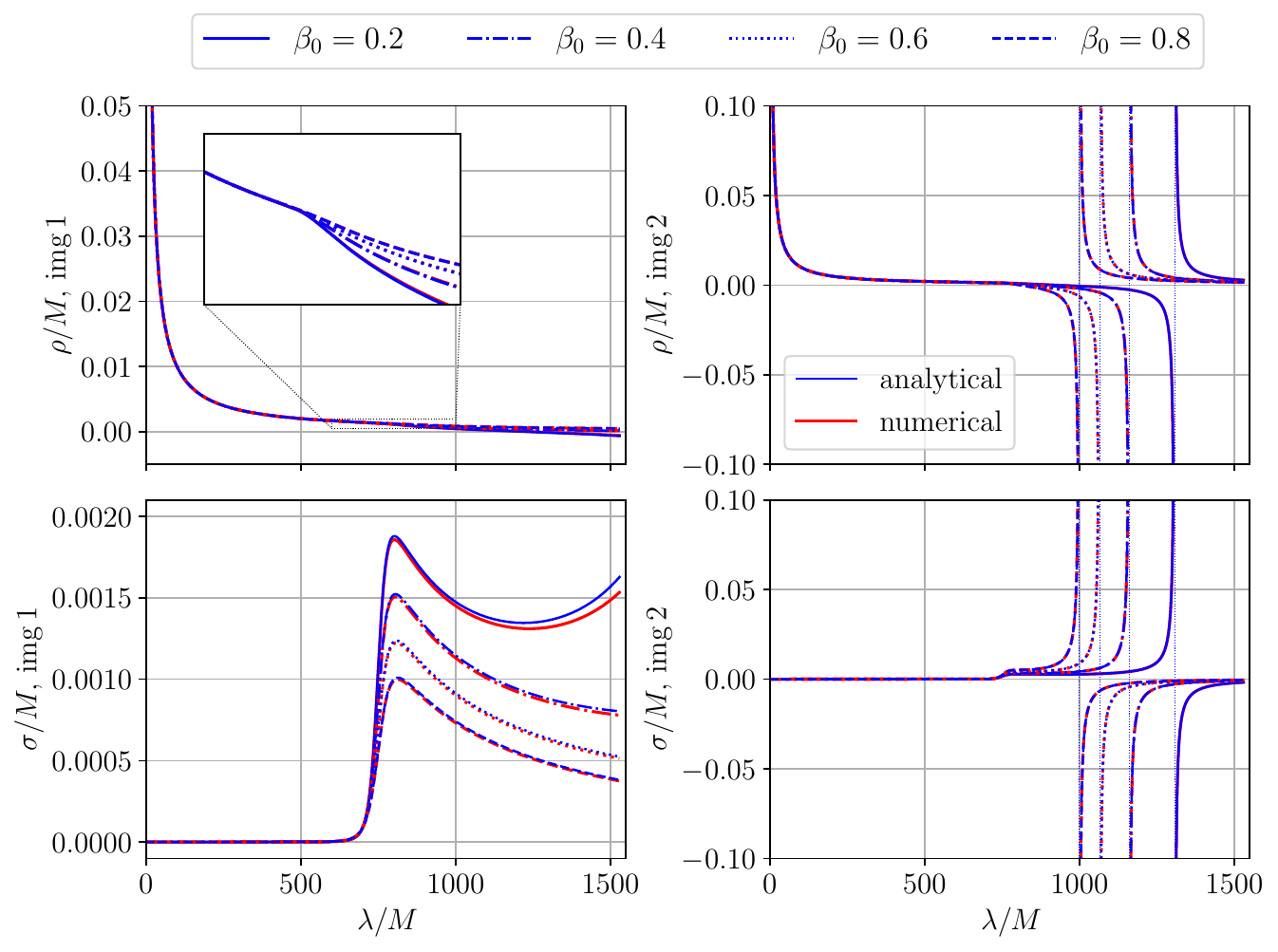"}
\caption{The evolution of the spin coefficients $\rho$ and $\sigma$. Same with Fig.\,\ref{fig:geodesic_deviation_vector}, the blue and red curves are obtained from the analytical solutions and corresponding numerical solutions to Eq.\,(\ref{delta-pm-equation}). The vertical dotted lines in the right panels represent the caustic points. The corresponding values are listed in Table \ref{tab:lensing-parameters}.}
\label{fig:rho_and_sigma}
\end{figure}

As shown in Eq.\,(\ref{evolution-area-axis-ratio}) and its surrounding context, we can define the cross-sectional area ($\mathcal{A}$) and axis ratio ($\mathcal{R}$) based on the deviation vector. Fig.\,\ref{fig:magnification_and_axis_ratio} illustrates their evolution along the photon/graviton path. For image 1 (shown in the left panels), both the area and axis ratio are always positive, but they approach zero at the caustic point in image 2 (shown in the right panels). The relative rate of change of $\mathcal{A}$ and $\mathcal{R}$ are defined as the optical scalars, $\rho$ and $\sigma$, respectively, which are shown in Fig.\,\ref{fig:rho_and_sigma}. As the photon/graviton approaches the wave source (point-like as we have assumed), all geodesics in congruence focus on a single point, subsequently leading to $\rho$ becoming infinite. Initially, $\tilde{\xi}_+$ and $\tilde{\xi}_-$ are set with the same initial conditions, such that $\sigma$ remains zero until the photon/graviton approaches the lens object. Notably, in image 1, both optical scalars remain continuous, while in image 2, a discontinuity arises due to the caustic point.

The weak deflection and thin-lens approximation introduces errors in the analytical solutions. For example, slight deviations can be seen between the slopes of the numerical and analytical solutions for $\tilde{\xi}_{\pm}$ (see Fig.\,\ref{fig:geodesic_deviation_vector}). These errors are amplified when calculating $\sigma$, particularly as the photon/graviton approaches the observer, as shown in the bottom-left panel of Fig.\,\ref{fig:rho_and_sigma}.  It is important to note that the parameters listed in Table \ref{tab:lensing-parameters} deviate significantly from those in realistic systems. A typical lens object, with mass $M\sim10^8M_{\odot}$ and redshift $z\sim1$, has an angular separation of the image and lens of approximately $\theta\sim1''$. We can estimate the distance and impact parameter for such a system as $D_L\approx4.3\times10^3\,\mathrm{Mpc}\approx9.0\times10^{14}M$ and $b\sim4.8\times10^{-6}D_L$. This estimation shows that the impact parameter is much larger than the Schwarzschild radius and that $D_L$ is much greater than the impact distance, validating the weak deflection and thin-lens approximations.

\subsection{\label{sec:magnification}The magnification and axis ratio of the lensing images}

In the last subsection, we explored the behaviors of the geodesic deviation vector, cross-sectional area/axis ratio, and spin coefficients along the photon/graviton trajectories. In this subsection, we turn to the observational imprints of these evolutions. The magnification of a lensing image is defined as the ratio of the lensed to the unlensed energy flux at the observer. In geometric optics, the energy flux of light and GWs is covariantly conserved within the congruence, making it proportional to $1/\mathcal{A}$. Therefore, the magnification can be written as
\begin{equation}
\label{magnification-definition}
\mu=\frac{\widehat{\mathcal{A}}(\lambda_o)}{\mathcal{A}(\lambda_o)}.
\end{equation}
where $\widehat{\mathcal{A}}$ represents the cross-sectional area of a congruence propagating freely along the unlensed path (OS, see Fig.\,\ref{fig:lensing}). By setting the initial conditions $\widehat{\mathcal{A}}(\lambda=0)=0$ and $\bm{\mathcal{D}}\widehat{\mathcal{A}}(\lambda=0)=1$, the solution is
\begin{equation}
\label{area-unlensed}
\widehat{\mathcal{A}}(\lambda)=\frac{1}{\mathrm{OS}^2},\quad\text{with}\quad
\text{OS}=D_S\sqrt{1+\tan^2\beta}
=\frac{r_o}{\sqrt{1+\tan^2\beta}}\left[1+\sqrt{1-(1+\tan^2\beta)\left(1-\frac{r_s^2}{r_o^2}\right)}\right].
\end{equation}
We recall that $r_s$ and $r_o$ are the radial Schwarzschild coordinates of the source and observer, which is related with $D_L$ and $D_S$ by
\begin{equation}
\label{ro-rs}
r_o=D_{L},\quad\text{and},\quad r_s=D_S^2(1+\tan^2\beta)-2D_LD_S+D_L^2
\end{equation}
The area of the lensed congruence is computed using the Dyson-like series in Eq.\,(\ref{In-int}). At the observer position, we obtain the following expressions for the first four terms in this series,
\begin{equation}
\label{lambda-o-solution}
\mathcal{I}_0(\lambda_o)=\lambda_o\approx\frac{M}{u_m}
\left(\frac{1}{x_s}+\frac{1}{x_o}\right)
\left[1-\frac{1}{2}x_sx_o\left(1-\frac{4u_m}{x_s+x_o}\right)\right],
\end{equation}
\begin{equation}
\label{I1-lambda-o}
\mathcal{I}_{1}(\lambda_o)
\approx\frac{2M}{x_sx_o}
\left\{2+\left(\frac{45\pi}{16}-4\right)u_m
+\left[\left(\frac{192}{5}-\frac{135\pi}{16}\right)u_m^2
+2u_m(x_s+x_o)-(x_s^2-x_sx_o-x_o^2)\right]\right\},
\end{equation}
\begin{equation}
\label{I2-lambda-o}
\mathcal{I}_2(\lambda_o)\approx\frac{M}{x_sx_o}
\left\{\frac{15\pi}{8}u_m
+u_m\left[\left(48-\frac{45\pi}{8}\right)u_m-4(x_s+x_o)\right]\right\},
\end{equation}
and
\begin{equation}
\label{I3-lambda-o}
\mathcal{I}_3(\lambda_o)
\approx\frac{M}{x_sx_o}\frac{16}{5}u_m^2.
\end{equation}
Under the weak deflection and thin-lens approximations, $u_m\ll1$, and $x_{s,o}\equiv r_m/r_{s,o}\ll1$ with the same order of magnitude as $u_m$, we expand these expressions in terms of small parameters $u_m$, $x_s$, and $x_o$, up to the zeroth order, as shown in Eqs.\,(\ref{lambda-o-solution}\,-\,\ref{I3-lambda-o}). By inserting these into Eq.\,(\ref{GDV-Dyson}), and then into the definition in Eq.\,\eqref{area-ratio-definition}, we can compute the area $\mathcal{A}(\lambda_o)$. Finally, combining this with the unlensed area in Eq.\,\eqref{area-unlensed} and applying Eq.\,\eqref{magnification-definition}, we obtain the magnification.

Next, we transform the parameters used here into the commonly employed lensing parameters. This is done as follows: (1) Transform $u_m$ into the impact parameter $b$ via Eq.\,(\ref{b-um-transformation}), where $b=\pm D_{L}\sin\theta/M$ for image 1 and 2, respectively. (2) Transform $x_s$ and $x_o$ into $\beta$, $D_L$, and $D_{LS}$ using their definition and Eq.\,(\ref{ro-rs}). (3) Substitute the relation between $\beta$ and $\theta$ from lens equation in Eq.\,\eqref{lens-equation-img1}. (4) In the weak deflection and thin lens limit, we expand the image position as
\begin{equation}
\label{theta-expansion}
\theta=\theta_E\left[\theta_0+\theta_1\epsilon+\theta_2\epsilon^2+\mathcal{O}(\epsilon^3)\right],
\end{equation}
where the bookkeeper $\epsilon$ is defined as,
\begin{equation}
\label{epsilon-definition}
\epsilon\sim\frac{\theta_E}{4\mathcal{D}},
\quad\text{with}\quad
\mathcal{D}\equiv\frac{D_{LS}}{D_S}.
\end{equation}
After these transformations, the magnification is expanded in powers of $\epsilon$ as
\begin{equation}
\label{magnification}
\begin{aligned}
\mu&=\frac{\theta_0^4}{\theta_0^4-1}\mp\frac{15\pi\theta_0^3}{16(1+\theta_0^2)^3}\epsilon
+\frac{\theta_0^2}{384(1-\theta_0^2)^2(1+\theta_0^2)^5}
\Bigg\{-1024(3-2\mathcal{D}^2)\\
&\quad+1024(9+12\mathcal{D}-14\mathcal{D}^2)\theta_0^2
+[4096(3-\mathcal{D}^2)-2025\pi^2]\theta_0^4
-\left[2048(6+15\mathcal{D}-17\mathcal{D}^2)-2025\pi^2\right]\theta_0^6\\
&\quad-1024(9+6\mathcal{D}-8\mathcal{D}^2)\theta_0^8
+1024\mathcal{D}(3+18\mathcal{D}-20\mathcal{D}^2)\theta_0^{10}
+6144\mathcal{D}(1-\mathcal{D})\theta_0^{12}\Bigg\}\epsilon^2+\mathcal{O}(\epsilon^3),
\end{aligned}
\end{equation}
The leading-order result is well-documented in standard textbooks, such as \cite{Schneider1992}. When $\theta=\theta_{E}$, the Einstein angle, and $\theta_0=1$, the magnification diverges to infinity. These positions, where the magnification becomes infinite, define the critical curve. When the image lies outside the critical curve, the magnification is positive, corresponding to an upright image with positive parity. Conversely, inside the critical curve, the magnification is negative, indicating an inverted image with negative parity, as summarized in Table \ref{tab:lensing-parameters}. In the second-order terms, the $\pm$ sign distinguishes between the two images: the upper sign for image 1 and the lower for image 2. The higher-order corrections have been derived by Ref.\,\cite{Keeton2005} using the Jacobian matrix formalism.

Let us compare our result (\ref{magnification}) with the previous work \cite{Keeton2005} [Eq.\,(76-79), with $A_1=4$, $A_2=15\pi/4$, and $A_3=128/3$] and highlight the differences. In \cite{Keeton2005}, the magnification is defined using the Jacobian matrix,
\begin{equation}
\label{magnification-Jocabian-matrix}
\mu=\left|\frac{\sin\beta}{\sin\theta}\frac{d\beta}{d\theta}\right|^{-1},
\end{equation}
which is consistent with Eq.\,(\ref{magnification}) at the leading order, $\mathcal{O}(\epsilon^0)$. The inconsistency appears at the first and second-order corrections, due to some missing corrections when Ref.\,\cite{Keeton2005} derives the higher-order corrections. 

(1): It is noted that the images formed at the same and opposite sides to the source correspond to positive and negative angular momentum. Therefore the lens equations for images 1 and 2 are slightly different if we require the impact parameters to be positive. This is shown in Eqs.\,(\ref{lens-equation-img1},\,\ref{lens-equation-img2}). However, Ref.\,\cite{Keeton2005} defined the impact parameter as $b\equiv L/ME$, rather than $b\equiv |L|/ME$, thus missing an extra $\pm$ symbol in the first-order correction. 

(2): As mentioned before, the plane perpendicular to the optical axis and passing through the lens object is usually referred to as the lens plane. As shown in Fig.\,\ref{fig:lensing}, the deflection point (marked by A) is not at such plane, unless $\mathrm{SN}=0$ and $D_{L}=D_{LS}$. However, Ref.\,\cite{Keeton2005} assumed that point A always lies on the lens plane, which leads to inconsistency in the second order. The above two corrections are raised from the lens equation derived by Ref.\,\cite{Keeton2005}, which is incompatible with high-order approximations of weak deflection limit. Once a more accurate lens equation is given, one can still calculate the magnification from the Jacobian matrix (\ref{magnification-Jocabian-matrix}).

(3): The definition in Eq.\,(\ref{magnification-Jocabian-matrix}) ignored the slight difference between paths SAO and SO, shown in Fig.\,\ref{fig:lensing}. In the unlensed case, the signal detected by the observer is emitted from the source in the direction of SO, while the lensed signal is emitted in the direction of SA. To calculate such corrections, we denote the energy flux along SO and SA as $\mathcal{F}_{\mathrm{SO}}$ and $\mathcal{F}_{\mathrm{SA}}$, respectively. With the help of SZ, which is perpendicular to SN and shown as the red line in Fig.\,\ref{fig:lensing}, the ratio $\mathcal{F}_{\mathrm{SA}}/\mathcal{F}_{\mathrm{SO}}$ is calculated as 
\begin{equation}
\label{F-SA-F-SO}
\frac{\mathcal{F}_{\mathrm{SA}}}{\mathcal{F}_{\mathrm{SO}}}
=\frac{\cos\angle\mathrm{ZSA}}{\cos\angle\mathrm{ZSO}}
=\frac{\cos(\alpha\mp\theta)}{\cos\beta},
\end{equation}
where we have used $\angle\mathrm{ZSO}=\beta$ and $\angle\mathrm{ZSA}=\pi/2-\angle\mathrm{ASI}=\pi/2-(\pi-\alpha-\angle\mathrm{SIA})=\alpha\mp\theta$, with $\mp$ corresponding to the image 1 and image 2. Considering corrections (1) and (2), and multiplying the factor (\ref{F-SA-F-SO}) onto Eq.\,(\ref{magnification-Jocabian-matrix}), one obtains the improved magnification, which is consistent with Eq.\,(\ref{magnification}).

Another direct observational quantity is the axis ratio of the images, which describes the shape distortion of the background wave source. Similar to the derivation of magnification, the axis ratio $\mathcal{R}(\lambda_o)$ is given by,
\begin{equation}
\label{axis-ratio}
\begin{aligned}
\mathcal{R}&=\frac{\theta_0^2-1}{\theta_0^2+1}
\mp\frac{15\pi(\theta_0^2-1)(1+3\theta_0^2)}{16\theta_0(1+\theta_0^2)^3}\epsilon
+\frac{(\theta_0^2-1)}{768\theta_0^2(1+\theta_0^2)^5}
\Bigg\{-9(2048-225\pi^2)\\
&\quad-[4096(18+3\mathcal{D}-4\mathcal{D}^2)-8100\pi^2]\theta_0^2
-3[4096(6+\mathcal{D}-\mathcal{D}^2)-3375\pi^2]\theta_0^4\\
&\quad-8193\mathcal{D}^2\theta_0^6
+6144(3-2\mathcal{D}+2\mathcal{D}^2)\theta_0^8
-4096\mathcal{D}(3-4\mathcal{D})\theta_0^{10}
\Bigg\}\epsilon^2+\mathcal{O}(\epsilon^3).
\end{aligned}
\end{equation}
The axis ratio is zero when the lensing image is located at the critical curve, with $\theta_0=1$. This means the extended source is distorted into a line segment, with zero area and infinite amplitude. Similar to the magnification, the axis ratio is positive/negative for the images located outside/inside the critical curve, representing whether the image is upright (positive parity) or inverted (negative parity). The numerical results of $\mathcal{R}$ are listed in Table \ref{tab:lensing-parameters}.

In the leading order of weak deflection limit, the magnification and axis ratio are related to the convergence ($\kappa$) and shear ($\gamma_1$ and $\gamma_2$) by 
\begin{equation}
\mu=\frac{1}{(1-\kappa)^2-\gamma^2},\quad\text{and}\quad\mathcal{R}=\frac{1-\kappa-\gamma}{1-\kappa+\gamma},
\end{equation}
with $\gamma=\sqrt{\gamma_1^2+\gamma_2^2}$. The convergence depends on the lens mass density within the geodesic congruence. The shear $\gamma_1$ characterizes the stretching of the lensing images and $\gamma_2$ characterizes its orientation. For the Schwarzschild lens, which is vacuum and axisymmetric, $\kappa=0$, $\gamma_2=0$, and $\gamma=\gamma_1$. Through the Jacobian matrix \cite{Schneider1992}, the shear of the point-mass lens is calculated as $\gamma=1/\theta_0^2$ at the leading order of weak deflection approximation, resulting in $\mu=\theta_0^4/(\theta_0^4-1)$ and $\mathcal{R}=(\theta_0^2-1)/(\theta_0^2+1)$, as shown in the leading terms of Eqs.\,(\ref{magnification}) and (\ref{axis-ratio}). However, higher-order correction has been rarely studied. As calculated by Ref.\,\cite{Crisnejo2018}, $\kappa$ and $\gamma_2$ are expected to be nonzero at the subleading order. More importantly, besides convergence and shear, another optical scalar, rotation, should be defined to represent the anti-symmetric component of the Jacobian matrix. We are developing this calculation, which goes beyond the scope of this work and will be investigated in our future work.

\section{\label{sec:conclusion}Conclusion}

In this work, we revisited the Schwarzschild lensing of light or GWs from geodesic deviation. The detected lensed signal consists of a congruence of light/GW rays, and their collective behavior is described by the GDE. To solve it, we projected the GDE onto the NP tetrad constructed along the null geodesics and got a pair of decoupled equations for $\xi_{+}$ and $\xi_{-}$ (\ref{delta-pm-equation}) for real Weyls scalar $\Psi_0$. Similar to the Dyson series, we presented the solution to the transverse components $\bm{\xi}_{\pm}$ of the deviation vector in Eqs.\,(\ref{GDV-Dyson}) and (\ref{In-int}). In addition, the relationship between the deviation vector and the optical scalars, $\rho$ and $\sigma$, were presented in Eq.\,(\ref{GDE-1-NP-m}). The physical meaning of these two scalars is the relative rate of change of the congruence cross section and the axis ratio [see Eq.\,(\ref{rho-sigma})]. 

Subsequently, we considered the lensing process in the weak deflection limit, where the impact parameter is sufficiently large. The general signature is shown in Fig.\,\ref{fig:lensing}; two images are formed by Schwarzschild lensing. In this assumption, the Dyson-like series was calculated from Eq.\,(\ref{Dyson-int-x-1},\,\ref{Dyson-int-x-2}) analytically. We depicted such results in Fig.\,\ref{fig:Dyson_Series}, and compared them with the numerical solution for a group of parameters listed in Table\,\ref{tab:lensing-parameters}. We also found that $\mathcal{I}_0$ and $\mathcal{I}_1$ are dominant, and $\mathcal{I}_2,\,\mathcal{I}_3$ are close to zero, implying the series (\ref{GDV-Dyson}) is convergent in the weak deflection limit. Fig.\,\ref{fig:geodesic_deviation_vector} shows the evolution of the deviation vector. $\tilde{\xi}_{+}$ increases more rapidly after lensing and keeps positive. And $\tilde{\xi}_{-}$ increases more slowly or even decreases after lensing, meeting the caustic point, and finally forms an inverted image at the observer. Correspondingly, Figs.\,\ref{fig:magnification_and_axis_ratio} and \ref{fig:rho_and_sigma} show the cross-sectional area, axis ratio, and optical scalars, defined in Eqs.\,(\ref{area-ratio-definition},\,\ref{evolution-area-axis-ratio}). 

In subsection \ref{sec:magnification}, we reproduced the magnification from the geodesic deviation [see Eq.\,(\ref{magnification})], which is consistent with the one obtained using the Jacobian matrix in Ref.\,\cite{Keeton2005}. Meanwhile, we found some missing corrections in the previous calculation, e.g., the deflection point deviates from the lens plane and the lens equations are slightly different for the two images. Additionally, we presented the result of the axis ratio, equivalently, the shear, of lensing images up to the second-order weak deflection approximation [see Eq.\,(\ref{axis-ratio})].

This work provides a solid foundation for future studies. As discussed in Ref.\,\cite{Dolan2018b,Shipley2019}, the calculation of spin coefficients is indispensable when investigating the lensing effects beyond geometric optics, in which richer information on light and GW polarization is presented \cite{Harte2019a,Harte2019b,Cusin2020,Dalang2022,ZhaoLi2022,Kubota2024}. In this work, we calculated $\rho$ and $\sigma$ following the procedure in Ref.\,\cite{Dolan2018a} in the simplest case, the Schwarzschild lensing. Furthermore, the astrophysical lens objects are more complex than a Schwarzschild black hole \cite{Keeton2002}. Extending the computation to a more general lens model is beneficial for lensing observations and tests of gravitational theories in the future.

\begin{acknowledgments}
We would like to thank Takahiro Tanaka, Emanuel Gallo, Shaoqi Hou, Sam Dolan, Haofu Zhu, Xinyue Jiang, Donglin Gao, and the anonymous referee for their helpful discussions and comments. This work is supported by Strategic Priority Research Program of the Chinese Academy of Science (Grant No. XDB0550300), the National Key R\&D Program of China Grant No.\,2022YFC2200100 and 2021YFC2203102, NSFC No.\,12273035 and 12325301 the Fundamental Research Funds for the Central Universities under Grant No. WK3440000004, and the science research grants from the China Manned Space Project with No.CMS-CSST-2021-B01, and China Scholarship Council, No.\,202306340128. X. G. acknowledges the fellowship of China National Postdoctoral Program for Innovative Talents (Grant No. BX20230104). T.L. is supported by NSFC No.\,12003008. T.Z. is supported in part by the National Key Research and Development Program of China under Grant No. 2020YFC2201503, the Zhejiang Provincial Natural Science Foundation of China under Grant Nos. LR21A050001 and LY20A050002, the National Natural Science Foundation of China under Grant No. 12275238.

\end{acknowledgments}

\appendix

\section{Derivation to Eq.\,(\ref{GDV-1}) and (\ref{GDV-2})}
\label{app:11_and_12}
From left-hand side of Eq.\,(\ref{GDV-1}), we have
\begin{equation}
\label{GDV-1-derivation-left}
\begin{aligned}
m_{\nu}\left(\xi^\alpha\nabla_{\alpha}k^{\nu}\right)
&=m_{\nu}\left[\xi^{(a)}e^{\alpha}_{(a)}\right]\nabla_{\alpha}k^{\nu}
=\xi^{(a)}\left\{e^{\alpha}_{(a)}(\nabla_{\alpha}k_{\nu})m^{\nu}\right\}\\
&=\xi^{(3)}\left\{e^{\alpha}_{(3)}(\nabla_{\alpha}k_{\nu})m^{\nu}\right\}
+\xi^{(4)}\left\{e^{\alpha}_{(4)}(\nabla_{\alpha}k_{\nu})m^{\nu}\right\}\\
&=\xi^{m}\Big[
m^{\alpha}(\nabla_{\alpha}k_{\nu})m^{\nu}\Big]
+\xi^{\bar{m}}\Big[
m^{\alpha}(\nabla_{\alpha}k_{\nu})\bar{m}^{\nu}\Big]
=\sigma\xi^{m}+\rho\xi^{\bar{m}}.
\end{aligned}
\end{equation}
And from the right-hand side of Eq.\,(\ref{GDV-1}), we have
\begin{equation}
\label{GDV-1-derivation-right}
\begin{aligned}
m_{\nu}\left(k^{\alpha}\nabla_{\alpha}\xi^{\nu}\right)
&=m_{\nu}\left[k^{\alpha}(\nabla_{\alpha}\xi^{\nu})\right]
=m_{\nu}\left[k^{\alpha}(\partial_{\alpha}\xi^{\nu}+\Gamma_{\alpha\lambda}^{\nu}\xi^{\lambda})\right]\\
&=m_{\nu}\left\{
k^{\alpha}\partial_{\alpha}\left[\xi^{(a)}e_{(a)}^{\nu}\right]
+k^{\alpha}\Gamma_{\alpha\lambda}^{\nu}
\left[\xi^{(a)}e_{(a)}^{\lambda}\right]\right\}\\
&=m_{\nu}\left\{
\left[k^{\alpha}\partial_{\alpha}\xi^{(a)}\right]e_{(a)}^{\nu}
+\xi^{(a)}k^{\alpha}\left[\partial_{\alpha}e_{(a)}^{\nu}+\Gamma_{\alpha\lambda}^{\nu}e_{(a)}^{\lambda}\right]\right\}\\
&=m_{\nu}\left\{
\left[\bm{\mathcal{D}}\xi^{(a)}\right]e_{(a)}^{\nu}
+\xi^{(a)}k^{\alpha}\nabla_{\alpha}e_{(a)}^{\nu}\right\}
=m_{\nu}\left[\bm{\mathcal{D}}\xi^{(a)}\right]e_{(a)}^{\nu}
=\bm{\mathcal{D}}\xi^{\bar{m}}.
\end{aligned}
\end{equation}
Combining them gives $\bm{\mathcal{D}}\xi^{\bar{m}}=\sigma\xi^{m}+\rho\xi^{\bar{m}}$, with its complex conjugate being Eq.\,(\ref{GDE-1-NP-m}). Similar with the derivation in Eq.\,(\ref{GDV-1-derivation-right}), the second derivative to $\xi^{\nu}$ is
\begin{equation}
\label{GDV-2-derivation-left}
m_{\nu}\frac{D^2\xi^{\nu}}{D\lambda^2}
=m_{\nu}\frac{Dz^{\nu}}{D\lambda}
=m_{\nu}k^{\alpha}\nabla_{\alpha}z^{\nu}
=\bm{\mathcal{D}}z^{\bar{m}}
=\bm{\mathcal{D}}\left(m_{\nu}z^{\nu}\right)
=\bm{\mathcal{D}}\left[m_{\nu}\left(k^{\alpha}\nabla_{\alpha}\xi^{\nu}\right)\right]
=\bm{\mathcal{D}}^2\xi^{\bar{m}}.
\end{equation}
And then the left-hand side of Eq.\,(\ref{GDV-2}) is 
\begin{equation}
\label{GDV-2-derivation-right}
-m_{\nu}R^{\nu}_{\rho\alpha\beta}
k^{\rho}k^{\beta}\xi^\alpha
=-C_{\nu\rho\alpha\beta}
m^{\nu}k^{\rho}e_{(a)}^\alpha k^{\beta}\xi^{(a)}
=-C_{\nu\rho\alpha\beta}
m^{\nu}k^{\rho}m^\alpha k^{\beta}\xi^{m}
=-C_{mkmk}\xi^{m}=\Psi_0\xi^m,
\end{equation}
where $C_{\nu\rho\alpha\beta}=R_{\nu\rho\alpha\beta}$ is the Weyl tensor in vacuum, and we have used $\xi^{n}=0$ and $R_{mk\bar{m}k}=0$. The latter can be proved as follows. Firstly, from $C_{mk\bar{m}k}=\overline{C_{\bar{m}kmk}}=\overline{C_{mk\bar{m}k}}$, we find $C_{mk\bar{m}k}$ is real. And then
\begin{equation}
C_{mk\bar{m}k}=\frac{1}{2}(C_{mk\bar{m}k}+C_{\bar{m}kmk})
=\frac{1}{2}C_{\nu\rho\alpha\beta}
(m^{\nu}\bar{m}^{\alpha}+\bar{m}^{\nu}m^{\alpha})k^{\rho}k^{\beta}
=\frac{1}{2}C_{\nu\rho\alpha\beta}
(g^{\nu\alpha}+k^{\nu}n^{\alpha}+k^{\alpha}n^{\nu})k^{\rho}k^{\beta}=0,
\end{equation}
where the background metric is $g^{\alpha\beta}=-k^{\alpha}n^{\beta}-k^{\beta}n^{\alpha}+m^{\alpha}\bar{m}^{\beta}+m^{\beta}\bar{m}^{\alpha}$. Combining Eqs.\,(\ref{GDV-2-derivation-left}) and (\ref{GDV-2-derivation-right}), we get $\bm{\mathcal{D}}^2\xi^{\bar{m}}=\Psi_0\xi^m$, with the complex conjugate being Eq.\,(\ref{GDE-2-NP-m}).

\section{\label{app:lensing}Schwarzschild lensing}
One defines $u\equiv M/r$, and then the photon/graviton trajectory is dictated by \cite{Chandrasekhar1983}
\begin{equation}
\label{u-varphi-equation-1}
\left(\frac{du}{d\varphi}\right)^2
=\frac{1}{b^2}-u^2(1-2u)\equiv f(u),
\end{equation}
The impact parameter is $b=|L|/M$. When $b>3\sqrt{3}$, the function $f(u)$ possesses two different zero points, denoted by $0<u_2<u_3$, the allowed orbits should be located at interval $[0,u_2]$ and $[u_3,+\infty]$, where $f(u)>0$. These first kinds of orbits are open, in which the massless particle moves between an infinite region and a pericenter, corresponding to the gravitational lensing situation. The radial range of allowed motion is $r_0<r<\max\{r_o,r_s\}$, or equivalently, $\min\{u_o,u_s\}<u<u_m\equiv M/r_m$. $r_o$ and $r_s$ denote the radial coordinates of the wave source and observer. correspondingly, $u_o\equiv M/r_o$ and $u_s\equiv M/r_s$.

In this work, we only focus on the weak deflection limit and thin lens approximation, in which the impact distance ($Mb$) is much larger than the gravitational radius of the lens black hole, and the distances from emitter/observer to lens object are very larger than the impact distance. Using a bookkeeper, $\epsilon$, the order of magnitude impact parameter, source/observer distance are 
\begin{equation} u_m\sim\frac{1}{b}\sim\epsilon,\quad
\frac{r_m}{r_{s}},\frac{r_m}{r_{o}}\sim\epsilon.
\end{equation}
The following calculations are based on the Taylor expansion by the bookkeeper $\epsilon$. Additionally, the explicit relationship between $u_m$ and $b$ is given by $f(u_m)=0$, 
\begin{equation}
\label{b-um-transformation}
b=\frac{1}{u_m}\frac{1}{\sqrt{1-2u_m}}
=\frac{1}{u_m}\left[1+u_m
+\frac{3}{2}u_m^2+\frac{5}{2}u_m^3+\mathcal{O}(u_m^4)\right].
\end{equation}

Following the standard steps, the trajectory equation $u=u(\varphi)$ is
\begin{equation}
\label{u-solution}
u=u_m\cos\varphi
+u_m^2\left(1-\cos\varphi+\sin^2\varphi\right)
+u_m^3\left[\left(2-\frac{3}{4}\sin^2\varphi\right)\cos\varphi-2(1+\sin^2\varphi)+\frac{15}{4}\varphi\sin\varphi\right]
+\mathcal{O}(u_m^3),
\end{equation}
Up to order $\mathcal{O}(u_m^3)$. The deflection angle, $\alpha$, is obtained from the asymptotic behavior of solution (\ref{u-solution}), given by
\begin{equation}
\label{deflection-angle}
\alpha=4u_m+\left(-4+\frac{15}{4}\pi\right)u_m^2+\left(\frac{122}{3}-\frac{15\pi}{2}\right)u_m^3.
\end{equation}

After giving the deflection angle (\ref{deflection-angle}), we re-drive the lens equation in this appendix. We first investigate image 1 shown in Fig.\,\ref{fig:img1}, in which the source and image lie on the same side of the optical axis. In $\triangle$ASI, applying the sine theorem gives,
\begin{equation}
\label{B-6}
\frac{\text{SI}}{\sin\alpha}=\frac{\text{AI}}{\sin\angle\text{ASI}}\quad\Rightarrow\quad
\frac{\text{NI}-\text{SN}}{\sin\alpha}=\frac{\text{OI}-\text{OA}}{\sin[\pi/2-(\alpha-\theta)]}
=\frac{\text{OI}-\text{OA}}{\cos(\alpha-\theta)},
\end{equation}
where we use $\angle\text{ASI}=\pi-\alpha-\angle\text{AIS}=\pi-\alpha-(\pi/2-\theta)=\pi/2-(\alpha-\theta)$
and then
\begin{equation}
\left(\frac{\text{NI}}{\text{OI}}
-\frac{\text{SN}}{\text{OI}}\right)
\cos(\alpha-\theta)
=\left(1-\frac{\text{OA}}{\text{OI}}\right)\sin\alpha,
\end{equation}
or, equivalently
\begin{equation}
\left(\frac{D_{S}\tan\theta-D_S\tan\beta}{D_S\sec\theta}
\right)\cos(\alpha-\theta)
=\left(1-\frac{\text{OA}}{D_S\sec\theta}\right)\sin\alpha.
\end{equation}
Then the length of line OA can be expressed from the $\triangle$OAL by sine theorem,
\begin{equation}
\frac{\text{OL}}{\sin\angle\text{OAL}}=\frac{\text{OA}}{\sin\angle\text{OLA}}.
\end{equation}
Here, the line AL divides $\angle$OAS equally, such that the above equation becomes
\begin{equation}
\frac{D_L}{\sin(\pi-\alpha)/2}=\frac{\text{OA}}{\sin[\pi-\theta-(\pi-\alpha)/2]}.
\end{equation}
Using the expression of OA in terms of $D_L$, $\theta$, and $\alpha$,
\begin{equation}
\text{OA}=D_L\frac{\cos(\alpha/2-\theta)}{\cos(\alpha/2)}.
\end{equation}
We finally arrived at
\begin{equation}
\label{lens-equation-img1}
\left(\tan\theta-\tan\beta\right)\cos\theta\cos(\alpha-\theta)
=\sin\alpha-\frac{2D_L}{D_S}\cos\left(\frac{\alpha}{2}-\theta\right)\cos\theta\sin\frac{\alpha}{2}.
\end{equation}
By replacing $\theta$ by $-\theta$, and $\beta$ by $-\beta$ (equivalently, replace $\mathrm{NI}-\mathrm{SN}$ by $\mathrm{NI}+\mathrm{SN}$) in Eq.\,(\ref{lens-equation-img1}), we get the lens equation for image 2,
\begin{equation}
\label{lens-equation-img2}
-\left(\tan\theta-\tan\beta\right)\cos\theta\cos(\alpha+\theta)
=\sin\alpha-\frac{2D_L}{D_S}\cos\left(\frac{\alpha}{2}+\theta\right)\cos\theta\sin\frac{\alpha}{2},
\end{equation}
in which the source and image lie on the different sides of the optical axis.

To solve the lens equation (\ref{lens-equation-img1},\,\ref{lens-equation-img2}), we rescale the source position and image position as
\begin{equation}
\label{normalized-beta-theta}
\beta=\beta_0\theta_E,\quad
\theta=\theta_E\left[\theta_0+\theta_1\epsilon+\theta_2\epsilon^2+\mathcal{O}(\epsilon^3)\right],
\end{equation}
where the bookkeeper $\epsilon$ is defined in Eq.\,(\ref{epsilon-definition}). Then the lens equation at leading order is simplified as
\begin{equation}
\label{beta-0-solution}
\beta_0=\theta_0-\frac{1}{\theta_0},
\end{equation}
at the leading order, whose solution is
\begin{equation}
\label{theta-0-solution}
\theta_0=\frac{1}{2}\left[\beta_0\pm\sqrt{\beta_0^2+4}\right].
\end{equation}
In the second and third orders, the solutions to $\theta_1$ and $\theta_2$ are
\begin{equation}
\label{theta-1-solution}
\theta_1=\pm\frac{15\pi}{16(1+\theta_0^2)},
\end{equation}
and
\begin{equation}
\label{theta-2-solution}
\begin{aligned}
\theta_2&=\frac{1}{3\theta_0(1+\theta_0^2)^3}
\Bigg\{48-\frac{675}{256}\pi^2-16\mathcal{D}^2
+\left[8(9-3\mathcal{D}+2\mathcal{D}^2)
-\frac{675}{128}\pi^2\right]\theta_0^2\\
&\qquad\qquad\qquad+40\mathcal{D}^2\theta_0^4
-8(3-9\mathcal{D}+4\mathcal{D}^2)\theta_0^6
+8\mathcal{D}(6-5\mathcal{D})\theta_0^8\Bigg\}.
\end{aligned}
\end{equation}
The plus-minus symbol corresponds to the image 1 and 2, respectively. The leading-order result (\ref{theta-0-solution}) is consistent with Ref.\,\cite{Keeton2005}, but there is an extra plus-minus symbol in first-order correction (\ref{theta-1-solution}). And the second-order result (\ref{theta-2-solution}) is different, because the deflection point deviates from the lens plane.

\section{\label{app:lambda-sol}Solution to the radial motion}
In weak deflection limit, the radial potential reduces to 
\begin{equation}
U_{r}=\pm\sqrt{1-x^2}\left\{1-u_m\left(\frac{x^2}{1+x}\right)
-u_m^2\left[\frac{x^2(2+x)^2}{2(1+x)^2}\right]\right\}+\mathcal{O}(u_m^3).
\end{equation}
where $x\equiv u/u_m$, $u=M/r\leqslant u_m$, $u_m=M/r_m\ll1$, with $r_m$ is the radius of pericenter. When the particle moves toward the pericenter, the above equation is integrated by
\begin{equation}
\label{lambda_1}
\begin{aligned}
M^{-1}\lambda_{\rm I}=M^{-1}\int_{r_s}^{r}U_{r}^{-1}dr
&\approx-\frac{1}{u_m}\left(\frac{\nu}{x}-\frac{\nu_s}{x_s}\right)
-\left(\frac{\nu}{1+x}-\frac{\nu_s}{1+x_s}\right)\\
&\qquad+\frac{u_m}{2}\left[\frac{x\nu}{(1+x)^2}
-\frac{x_s\nu_s}{(1+x_s)^2}
+3\arcsin x-3\arcsin x_s\right],
\end{aligned}
\end{equation}
with initial condition $\lambda_s=0$. For simplicity, we define 
$\nu=\sqrt{1-x^2}$, and $\nu_s=\sqrt{1-x_s^2}$. When arriving the pericenter, $r=r_m$, where $x=1$, the affine parameter is
\begin{equation}
\lambda_m=\frac{1}{u_m}\frac{M}{x_s}
\left[1+x_s\left(u_m-\frac{x_s}{2}\right)\right],
\end{equation}
And then, when moving backward, the integration should be
\begin{equation}
\label{lambda_2}
\begin{aligned}
M^{-1}\lambda_{\rm II}&=M^{-1}\left\{\lambda_m+\int_{r_m}^{r}U_{r}^{-1}dr\right\}\\
&\approx\frac{1}{u_m}\left(\frac{\nu}{x}+\frac{\nu_s}{x_s}\right)
+\left(\frac{\nu}{1+x}+\frac{\nu_s}{1+x_s}\right)\\
&\qquad-\frac{u_m}{2}\left[\frac{x\nu}{(1+x)^2}
+\frac{x_s\nu_s}{(1+x_s)^2}
+3\arcsin x+3\arcsin x_s-3\pi\right].
\end{aligned}
\end{equation}
And the affine parameter at the observer position has been shown in Eq.\,(\ref{lambda-o-solution}).

\section{\label{app:analytical-In}The analytical expressions of Eqs.\,(\ref{Dyson-int-x-1}) and (\ref{Dyson-int-x-2})}
In this appendix, we present the analytical solution of Eqs.\,(\ref{Dyson-int-x-1}) and (\ref{Dyson-int-x-2}). The first three orders of the Dyson-like series are
\begin{equation}
\label{I1-analytic}
\begin{aligned}
M^{-1}\mathcal{I}_1&=\pm\left(1+\frac{x_s}{x}-\frac{2}{xx_s}\right)\nu
\pm\frac{u_m}{8xx_s(1+x)}\Big\{\Big[16(2+x)
-(61+61x-15x^2-7x^3)x_s\\
&\quad+8xx_s^2+7(1+x)x_s^2\Big]\nu
+15(3\nu\nu_s\pm xx_s)\arccos x\Big\}
\pm\frac{u_m^2}{40xx_s(1+x)^2}
\Big[-8(192+364x+167x^2)\\
&\quad+(835+1709x+1143x^2+326x^3+119x^4+42x^5)x_s
+(543+1086x+523x^2)x_s^2\\
&\quad+35x(1+x)x_s^3
+42(1+x)^2x_s^4\Big]\nu
\mp\frac{3u_m^2}{8xx_s(1+x)(1+x_s)}\Big\{
\pm15\left[3+2x+(2+x)x_s\right]\nu\nu_s\\
&\quad+(1+x)\left[-8x-(8+13x-4x^2)x_s
-(8+x-4x^2)x_s^2+4xx_s^3\right]\Big\}\arccos x\mp(x\leftrightarrow x_s),
\end{aligned}
\end{equation}
\begin{equation}
\label{I2-analytic}
\begin{aligned}
M^{-1}\mathcal{I}_2&=
\pm\frac{u_m}{8xx_s}\Big\{\left[16x
+(1-3x^2)x_s-8xx_s^2-3x_s^2\right]\nu
\pm3(5\nu\nu_s\mp xx_s)\arccos x\Big\}\\
&\quad\pm\frac{u_m^2}{8xx_s(1+x)(1+x_s)}
\Big[-2(96+112x+x^2-7x^3)
-(179+151x-60x^2-6x^3+5x^4)x_s\\
&\quad+(139+199x+47x^2-15x^3-5x^4)x_s^2
+(126+116x-22x^2-7x^3)x_s^3
-(5+15x+7x^2)x_s^4\\
&\quad-5(1+x)x_s^5\Big]\nu
+\frac{3u_m^2}{8xx_s(1+x)(1+x_s)}\Big\{
-5\left[3+5x+3x^2+(5+7x+3x^2)x_s+3(1+x)x_s^2\right]\nu\nu_s\\
&\quad\pm(1+x)\left[10x+(10+11x-5x^2)x_s
+(10-4x-5x^2)x_s^2-5xx_s^3\right]\Big\}\arccos x\mp(x\leftrightarrow x_s),
\end{aligned}
\end{equation}
and
\begin{equation}
\label{I3-analytic}
\begin{aligned}
M^{-1}\mathcal{I}_3&=
\pm\frac{u_m^2}{40xx_s}
\Big[-2(32+15x)-x(19-3x)x_s
+(107+15x^2)x_s^2+15xx_s^3+3x_s^4\Big]\nu\\
&\quad+\frac{3u_m^2(x+x_s)}{8xx_s}
(-5\nu\nu_s\mp2\pm xx_s)\arccos x\mp(x\leftrightarrow x_s).
\end{aligned}
\end{equation}
The upper sign corresponds to the case where $v_s<v<v_m$, and the lower one to $v_m<v<v_o$. $(x\leftrightarrow x_{s})$ represents exchanging all $x$ and $x_s$ in the formula.

\bibliographystyle{apsrev4-2}
\bibliography{reference}

\end{document}